\documentclass[11pt,a4paper]{article}

\usepackage{lineno}
\usepackage[hidelinks]{hyperref}

\modulolinenumbers[5]

\usepackage{authblk}

\usepackage{graphicx,amsfonts}
\hypersetup{
    colorlinks=false,
    linkcolor=blue,
    filecolor=magenta,
    urlcolor=blue,
}

\usepackage{bm}
\usepackage{amsmath,amssymb}
\usepackage{color}
\usepackage{comment}
\usepackage{ulem}

\newcommand\uv{\boldsymbol{u}}
\newcommand\utilde{\boldsymbol{\tilde{u}}}
\newcommand\fv{\boldsymbol{f}}
\newcommand\rhof{\rho_f}
\newcommand\rhop{\rho_p}
\newcommand\dd{\mathrm{d}}
\newcommand\uG{\boldsymbol{u_p}}
\newcommand\uP{\boldsymbol{u}}
\newcommand\omep{\boldsymbol{\omega_p}}
\newcommand\Ip{I_p}
\newcommand\rhs{\mathrm{\mathbf{rhs}}}

\newcommand{\revision}[2]{#2}
\newcommand{\revisions}[1]{}
\newcommand\MGV[1]{\textcolor{black}{#1}}
\newcommand\MGVN[1]{\textcolor{black}{#1}}
\newcommand\MU[1]{\textcolor{black}{#1}}

\usepackage[numbers]{natbib}

\begin{document}


\title{An efficient method for particle-resolved simulations \\
of neutrally buoyant spheres}


\author[1]{M. Garc{\'{\i}}a-Villalba}
\author[2]{B. Fuentes}
\affil[1]{Institute of Fluid Mechanics and Heat Transfer, TU Wien, 1060 Vienna, Austria}
\affil[2]{Universidad Carlos III de Madrid, Legan\'es 28911, Spain}

\author[3]{J. Du{\v{s}}ek}
\affil[3]{Universit{\'e} de Strasbourg, Fluid Mechanics Department, Institut ICube, France}
\author[1,4]{M. Moriche}
\author[4]{M. Uhlmann}
\affil[4]{Inst. for Hydromechanics, Karlsruhe Inst. of Technology, Karlsruhe 76131, Germany}

\date{}                     
\setcounter{Maxaffil}{0}




\maketitle 

\begin{abstract}
We present a simple modification of the
direct-forcing immersed boundary method (IBM) proposed by 
\citet{uhlmann:2005} in order to
enable it to be applied to particulate flows 
with solid-to-fluid
density ratios
around unity. 
%
%
\MU{%
  The main difference with respect to the original formulation lies in
  the particle velocity update which is performed directly after the
  preliminary velocity field has been computed in the absence of any
  IBM volume forcing term. In addition, we apply the forcing term to
  the entire space occupied by the immersed solid object (instead of
  to the vicinity of its interface only). 
}
%
\MU{%
  The present approach requires the evaluation of} 
integrals of the velocity 
field over the volume occupied by the solid particle,
which are evaluated efficiently as sums over the respective quantities
available at
particle-attached 
force points.
The resulting method can be
used seamlessly for density ratios down to $\rho_p/\rho_f>0.5$.
The new formulation has been validated using three configurations:
(i)  lateral migration of a neutrally buoyant circular particle in
two-dimensional Couette flow;
(ii)  the release from rest of a neutrally buoyant sphere in a free
stream;
(iii) the release of a particle in a free
stream after an initial phase in which it is
translationally fixed with an imposed angular velocity.
In all three test cases the present IBM formulation yields a very good 
agreement with the available reference data. 
Thus, the proposed approach is a
cost-efficient and accurate modification of the original method
which
allows for the simulation of fluid systems involving density-matched
solid particles.
\end{abstract}



\section{Introduction}

\MGV{Over} the last 20 years the numerical simulation of particle-laden flows has reached
a certain level of maturity. 
There is \MGV{now} a large number of particle-resolved techniques available in the literature \cite{maxey2017,wachs2019,uhlmann2022}.
Among them, the immersed boundary method (IBM) \cite{mittal2005,verzicco2023} is a popular \MGV{approach}
mainly due to the fact that the underlying computational grid 
does not need to be adapted over time, leading
to relatively simple algorithms.
The IBM was first developed by Peskin \cite{peskin1972,peskin2002} in the context of heart
flow simulations and since \MGV{then} multiple variations have been published and are in use
today for a variety of problems \cite{sotiropoulos2014,kim2019,griffith2020,mittal2022,arranz2022}.
One method that has found widespread use is the direct-forcing IBM developed by Uhlmann \cite{uhlmann:2005}.
As recently pointed out by Zhou and Balachandar \cite{zhou2021}, ``\MGV{[it]} 
 is very popular because it is easy to implement and flexible to simulate various problems''.
Over the years the method of Uhlmann has been modified and improved 
by several authors \cite{vanella2009,pinelli2010,kempe2012,breugem2012}.

Despite its success, the method of Uhlmann has \MGVN{a number of} limitations. 
\MGV{One of them} arises from the weak coupling between the Navier-Stokes equations
for the fluid phase and the Newton-Euler equations for the particle motion,
leading to a lower limit for the density ratio between solid and fluid phase of about 1.2. 
For lighter particles the computations are unstable which is unsurprising
due to the singularity of the \MGV{formulation} for the neutrally buoyant case.
\MGV{In order to} overcome this limitation there have been several proposals in the literature.
Kim and Choi \cite{kim2006} presented a method in which the particle equations
are integrated with an implicit method and then showed how the evaluation of the hydrodynamic
force at the new (unknown) time level can be avoided by using Lagrangian quantities.
This results in a non-iterative and non-singular formulation for the neutrally buoyant case.
However, to the best of our knowledge, this method has not  been applied yet to the motion of light particles.
A different approach was proposed by Schwarz et al. \cite{schwarz2015} by adding
a stabilizing term to both sides of the Newton-Euler equation. 
This term is denoted {\it virtual mass force} and allows for small density ratios to be 
computed without stability issues. 
A modification of \MGVN{Schwarz et al.}'s method has been reported by Tavanashad and Subramaniam \cite{tavanashad2020} for the case of larger volume fractions.

A further improvement of \MGVN{Schwarz et al.}'s 
method was reported by Tschisgale et al. \cite{tschisgale2017}.
These authors restrict the rigid-body assumption to a finite-thickness interface layer and
modify the coupling between fluid equations and particle equations, leading
to a non-interative semi-implicit coupling. 
The resulting method is non-singular and remains valid even in the limit of vanishing 
particle mass. 
The authors however report a weak oscillation of the particle velocity when the particle moves
free of external forces, a feature that they attribute to the numerical integration of
the virtual mass force.
In the present work, using some of the ideas of Tschisgale et al. \cite{tschisgale2017},
we propose and validate a new modification of the original method of Uhlmann \cite{uhlmann:2005}
for the computation of neutrally buoyant spheres.
\revision{
The new method does not require the addition of stabilizing virtual mass forces and 
is free of spurious oscillations. 
}
{As will be shown below, the new method is free of spurious oscillations.}

\MGVN{In order to validate the present technique, a new flow configuration is proposed,
in which the response of a neutrally buoyant particle involves non-trivial angular motion.
In this context, we have generated new high-fidelity reference data with the aid of 
a spectral-element method on a particle-attached conforming grid \cite{uhlmann2014}.}

The manuscript is organized as follows. 
First, the formulation of the problem,
the governing equations and a discussion of the fluid-particle coupling are provided
in \S \ref{sec:formulation}. 
This is followed by the numerical details and a complete description of the proposed method in \S 
\ref{sec:numerics}. 
We have performed three test cases for the validation of the methodology. 
The first is the migration of a neutrally buoyant disk in 2D Couette flow (\S \ref{sec:nb_couette}).
The second is a neutrally buoyant sphere released from rest in a free stream (\S \ref{sec:nb_free_stream}).
The third and final case is an initially rotating sphere released in a free-stream
(\S \ref{sec:rot}).
The manuscript ends with some final remarks and conclusions.

\section{Formulation of the problem} \label{sec:formulation}

The physical problem under consideration consists of the interaction of rigid particles
of constant density, $\rho_p$, with a surrounding viscous fluid of constant
kinematic viscosity, $\nu$, and constant density, $\rho_f$, in the incompressible flow limit.
For simplicity, we limit the discussion to a single particle, but the extension to several particles is trivial and has been discussed elsewhere \cite{uhlmann:2005}.
In this work we only consider a circular \MGV{particle} in 2D or a sphere in 3D. 
The extension of the method to particles of arbitrary shape does not present fundamental
problems but only the \MGV{added complexity} of tracking the orientation of the particles.
This \MGV{has already been} illustrated for \MGVN{related} methods in the literature
\cite{tschisgale2018,arranz:2018, moriche2021}.
%
%

\subsection{Fluid phase equations}

The governing equations for the fluid phase are the Navier-Stokes of the incompressible flow
for a Newtonian fluid
\begin{align}
&\nabla \cdot \uv = 0, \label{eq:continuity} \\
&\frac{\partial \uv}{\partial t} + (\uv \cdot \nabla)\uv = - \nabla p + \nu \nabla^2 \uv + \fv,\label{eq:momentum}
\end{align}
where $\uv=(u,v,w)$ is the fluid velocity, $p$ is the pressure divided by the fluid density, 
and $\fv$ is a \MGVN{specific} volume force term. 
As in the original method, the equations are enforced throughout the entire domain, $\Omega$,
comprising the fluid domain, $\Omega_f$ and the space occupied by the particle, $S$. 
These equations need to be supplemented with initial and suitable boundary conditions at the outer boundary. 
The volume force $\fv$ is \MGV{formulated in order to enforce} the no-slip boundary condition on the surface of the particle, as discussed in the following sections. 
In the original method, the forcing was restricted to a layer around the solid-fluid interface.
\MGV{
However, recent work has shown that good results are obtained  extending the forcing to the interior of the particle \cite{moriche2021}.
Here, we use the latter approach, as it was found beneficial 
 to treat neutrally buoyant particles  by other authors \cite{yu2007}. 
 }

\subsection{Solid phase equations}

The motion of the particle is governed by \MGVN{the Newton-Euler} equations for the motion of a rigid
body under the action of hydrodynamic and gravity forces
\begin{align}
\rhop V_c\frac{\dd\uG}{\dd t}&=\rhof\oint_{\partial S}\boldsymbol{\tau}\cdot\boldsymbol{n}\dd\sigma+(\rhop-\rhof)V_c\boldsymbol{g}, \label{eq:uG1}\\
I_p\frac{\dd\omep}{\dd t}&=\rhof\oint_{\partial S}\boldsymbol{r}\times(\boldsymbol{\tau}\cdot\boldsymbol{n})\dd\sigma \label{eq:omep1},
\end{align}
where $V_c$ is the volume of the particle, $\uG=(u_p,v_p,w_p)$ the velocity of the centre of mass \MGV{of the particle (located at $\boldsymbol{x_p}$), $\boldsymbol{g}$ the gravitational} acceleration, 
$\omep=(\omega_{px},\omega_{py},\omega_{pz})$ the angular velocity of the particle, 
$\boldsymbol{r}=\boldsymbol{x}-\boldsymbol{x_p}$ the position vector of any point in the body with respect to the centre of mass,
$\boldsymbol{n}$ the outward-pointing normal unit vector
at the particle surface
and
$\boldsymbol{\tau}=-p\boldsymbol{I}+\nu(\nabla \uv + \nabla \uv^T)$ is the hydrodynamic stress tensor
with $\boldsymbol{I}$ being the identity tensor.
In 3D $I_p$ is the moment of inertia with respect to any axis that passes through $G$, while in 2D $I_p$ is the moment of inertia with respect to the axis passing through $G$ and perpendicular to the motion.
For convenience, we also introduce the specific moment of inertia $\tilde{I}_p=I_p/\rhop$.
With respect to the hydrodynamic force and torque, following \cite{uhlmann:2005} we can write
\begin{align}
\oint_{\partial S}\boldsymbol{\tau}\cdot\boldsymbol{n}\dd\sigma & = -\int_S \fv\dd\boldsymbol{x}+\frac{\dd}{\dd t}\int_S \uv\dd\boldsymbol{x},\\
\oint_{\partial S}\boldsymbol{r}\times(\boldsymbol{\tau}\cdot\boldsymbol{n})\dd\sigma & =
 -\int_S (\boldsymbol{r}\times\fv)\dd\boldsymbol{x}+
\frac{\dd}{\dd t}\int_S (\boldsymbol{r}\times\uv)\dd\boldsymbol{x}.
\end{align}
In case of rigid body motion, it can be shown that the following equations hold
\begin{align}
\frac{\dd}{\dd t}\int_S \uv\dd\boldsymbol{x}& =  V_c \frac{\dd \uG}{\dd t},\\
\frac{\dd}{\dd t}\int_S (\boldsymbol{r}\times\uv)\dd\boldsymbol{x} & = \frac{\Ip}{\rhop}  \frac{\dd \omep}{\dd t}.
\end{align}
Then, we can rewrite equations (\ref{eq:uG1})-(\ref{eq:omep1}) as
\begin{align}
(\rhop-\rhof) V_c\frac{\dd\uG}{\dd t}&=-\rhof\int_S \fv\dd\boldsymbol{x}+(\rhop-\rhof)V_c\boldsymbol{g}, \label{eq:uG2}\\
\left(1-\frac{\rhof}{\rhop}\right)I_p\frac{\dd\omep}{\dd t}&=-\rhof\int_S (\boldsymbol{r}\times\fv)\dd\boldsymbol{x} \label{eq:omep2},
\end{align}

\subsection{Fluid-solid coupling} \label{sec:coupling}

The coupling between the particle motion and the fluid motion is realized 
via the volume force term $\fv$. 
We can solve for $\fv$ in eq. (\ref{eq:momentum}), and grouping together convective,
pressure and viscous terms into the variable $\rhs$, we may write
\begin{equation}
\fv=\frac{\partial \uv}{\partial t}-\rhs.\label{eq:coupling1}
\end{equation}
Following \cite{tschisgale2017}, we integrate equation (\ref{eq:coupling1})
over a time interval $[t_{n-1},t_n]$
\begin{equation}
\int_{t_{n-1}}^{t_n}\fv\dd t=\int_{t_{n-1}}^{t_n}\left(
\frac{\partial \uv}{\partial t}-\rhs
\right)\dd t
=\uv^n-\uv^{n-1}-\int_{t_{n-1}}^{t_n}\rhs\,\dd t,\label{eq:coupling2}
\end{equation}
where $\uv^n$ is the fluid velocity at the time instant $t_n$. 
Now we link the fluid and particle motions, by imposing that the fluid velocity
of any interior point
at the time instant $t_n$ coincides with the rigid body velocity 
\begin{equation}
\uv^n = \uG + \omep\times \boldsymbol{r}\quad \forall \boldsymbol{x}\in S.\label{eq:rigidbody1}
\end{equation}
We also introduce a preliminary velocity field obtained without the contribution of
the forcing term 
\begin{equation}
\utilde=\uv^{n-1}+\int_{t_{n-1}}^{t_n}\rhs\,\dd t.
\end{equation}
Then, we can rewrite eq. (\ref{eq:coupling2}) as
\begin{equation}
\int_{t_{n-1}}^{t_n}\fv\dd t=\uP^n-\utilde\quad \forall \boldsymbol{x}\in S.
\label{eq:coupling3}
\end{equation}

Note that, a similar analysis has been performed by \cite{tschisgale2017}. 
A key difference, however,
is that these authors only include the forcing term in a layer surrounding the particle
boundary in order to impose the no-slip boundary condition. 
In the present work, we also include the forcing term in the particle interior so that
\MGV{the fluid velocity in the interior of the particle is forced to follow the rigid-body motion.}

We can now also integrate the equations (\ref{eq:uG2})-(\ref{eq:omep2})
\MGVN{over the time interval $[t_{n-1},t_n]$}
\begin{align}
(\rhop-\rhof) V_c\left(\uG^n-\uG^{n-1}\right)&=-\rhof\int_S \int_{t_{n-1}}^{t_n}\fv\dd t\dd\boldsymbol{x}+(\rhop-\rhof)V_c\boldsymbol{g}\Delta t, \label{eq:uG3}\\
\left(1-\frac{\rhof}{\rhop}\right)I_p\left(\omep^n-\omep^{n-1}\right)&=-\rhof\int_S \int_{t_{n-1}}^{t_n} (\boldsymbol{r}\times\fv)\dd t\dd\boldsymbol{x} \label{eq:omep3},
\end{align}
where $\Delta t=t_n-t_{n-1}$. 
\MGVN{On} the right hand side of equations (\ref{eq:uG3}) and (\ref{eq:omep3}), the order of integration has been exchanged, as discussed in detail in Appendix A of Tschisgale et al.
\cite{tschisgale2017}.
Using equations (\ref{eq:rigidbody1}) and (\ref{eq:coupling3})
we can write
\begin{align}
\int_S \int_{t_{n-1}}^{t_n}\fv\dd t\dd\boldsymbol{x} & = \int_S(\uG^n+\omep^n\times\boldsymbol{r}-\utilde)\dd\boldsymbol{x}= V_c\uG^n-\int_S\utilde\dd\boldsymbol{x},
\label{eq:ibforce1}\\
\int_S \int_{t_{n-1}}^{t_n} (\boldsymbol{r}\times\fv)\dd t\dd\boldsymbol{x}&=
 \int_S\left[\boldsymbol{r}\times\left(
\uG^n+\omep^n\times\boldsymbol{r}-\utilde\right)\right]\dd\boldsymbol{x}=
\frac{\Ip}{\rhop}\omep^n-\int_S\boldsymbol{r}\times\utilde\dd\boldsymbol{x}.
\label{eq:ibtorque1}
\end{align}
Introducing equations (\ref{eq:ibforce1})-(\ref{eq:ibtorque1}) into equations (\ref{eq:uG3})-(\ref{eq:omep3}) and re-arranging terms we obtain
\begin{align}
\uG^n&=\left(1-\frac{\rhof}{\rhop}\right)\uG^{n-1}+\frac{1}{V_c}\frac{\rhof}{\rhop}\int_S\utilde\dd\boldsymbol{x}+\Delta t\left(1-\frac{\rhof}{\rhop}\right)\boldsymbol{g},\label{eq:uG4}\\
\omep^n&=\left(1-\frac{\rhof}{\rhop}\right)\omep^{n-1}+\frac{1}{\tilde{I}_p}\frac{\rhof}{\rhop}\int_S\boldsymbol{r}\times\utilde\dd\boldsymbol{x}\label{eq:omep4}.
\end{align}
The formulation (\ref{eq:uG4}-\ref{eq:omep4})
does not present a singularity at $\rhop/\rhof=1$, but it does in the limit
$\rhop/\rhof\rightarrow 0$. 
\MGV{
However,  the method is numerically unstable for $\rhop/\rhof< 0.5$.
This is because the coefficient that multiplies  $\uG^{n-1}$ in eq. (\ref{eq:uG4}) becomes
larger than 1 in absolute value for $\rhop/\rhof< 0.5$, yielding an unstable algorithm (see for example pages 13-14 in \cite{ascher2011}).
 }
For neutrally buoyant particles, $\rhop/\rhof=1$,  
expressions (\ref{eq:uG4}-\ref{eq:omep4}) simplify to 
\begin{align}
\uG^n&=\frac{1}{V_c}\int_S\utilde\dd\boldsymbol{x},\label{eq:uG5}\\
\omep^n&=\frac{1}{\tilde{I}_p}\frac{\rhof}{\rhop}\int_S\boldsymbol{r}\times\utilde\dd\boldsymbol{x}\label{eq:omep5},
\end{align}
\MGV{which has already been} proposed by \cite{sharma2005}. Note that in the neutrally buoyant limit, $\uG^n$ does not depend explicitly on $\uG^{n-1}$, 
\MGV{so that the motion of the particle and the motion of the virtual fluid occupying the position of the particle}
have to be compatible from the initial time.
\MGV{
In other words, the initial velocity field assigned in the interior of a particle needs to be compatible with the initial (linear and angular) 
particle velocity, according to rigid body motion, eq. (\ref{eq:rigidbody1}).
}
This is \MGVN{not the case} when $\rhof\neq\rhop$.

\section{Numerical methodology} \label{sec:numerics}

The numerical method to solve the governing equations is then very similar to the original method proposed by \cite{uhlmann:2005}. In particular we employ the same flow solver, using
second-order finite differences \MGV{on} a staggered grid and a semi-implicit 3-stage Runge-Kutta 
\MGVN{scheme}
where
the linear, viscous terms are treated implicitly and the convective, non-linear terms are
treated explicitly. 

We \MGVN{also} employ  separate discretizations for the Eulerian and Lagrangian quantities. 
We denote Eulerian quantities with lowercase letters and Lagrangian quantities with uppercase
letters.
For the Eulerian \MGVN{discretization}, we use a fixed, uniform, Cartesian grid $g_h$ of grid spacing $h$. 
The location of the grid points is denoted $\boldsymbol{x}_{ijk}^{\beta}$, 
\MGVN{where the superscript $\beta=1,2,3$ refers to the staggered grid associated with 
the velocity component $u_\beta$}.
For the Lagrangian grid, we evenly distribute $N_L$ points \MGV{throughout} the volume occupied by the
particle $S$. The \MGV{Lagrangian} locations are denoted $\boldsymbol{X}_\ell\in S$ \MGV{with} $1\leq\ell\leq N_L$.
To each point we associate a discrete volume $\Delta V_\ell$, such that the sum
of these volumes \MGV{equals} the total volume of the particle.
The transfer of quantities between Lagrangian and Eulerian
locations is based on the same regularized delta function
$\delta_h$ introduced by \cite{peskin2002} and defined by \cite{roma1999}, as in the original method
\cite{uhlmann:2005}, where additional details can be found.

It is now possible to describe the new proposed algorithm. 
This is done here in 3D, the reduction to 2D being \MGVN{straightforward}.
For the $k$th Runge-Kutta stage, first compute the preliminary fluid velocity $\bm{\tilde{u}}$
without accounting for the effect of the immersed boundary, viz.
\begin{equation}
\utilde = \uv^{k-1}+\Delta t\left(
2\alpha_k\nu\nabla^2\uv^{k-1}-2\alpha_k\nabla p^{k-1}
-\gamma_k[(\uv\cdot\nabla\uv]^{k-1}
-\xi_k[(\uv\cdot\nabla\uv]^{k-2}
\right),
\label{eq:utilde}
\end{equation}
where the coefficients $\alpha_k,\,\gamma_k,\,\xi_k\, (1\leq k\leq3)$ are taken from
\cite{rai1991}. 
Next, we transfer the preliminary velocity from the Eulerian to the Lagrangian grid
\begin{equation}
\tilde{U}_\beta(\boldsymbol{X}_\ell)=\sum_{i,j,k}\tilde{u}_\beta(\boldsymbol{x}_{ijk}^{\beta})
\delta_h\left(
\boldsymbol{x}_{ijk}^{\beta}-\boldsymbol{X}_\ell
\right)h^3\quad\forall\ell;1\leq\beta\leq3.
\label{eq:EuToLag}
\end{equation}

The next step in the original algorithm is the computation of the force volume term 
using the desired velocity at the Lagrangian points computed from the values of the previous stage $k-1$.
In the present algorithm we instead compute next the new velocity of the centre of mass of the particle, $\uG^{k}$,
and the new particle angular velocity, $\omep^k$, using equations (\ref{eq:uG4})-(\ref{eq:omep4}),
or for the neutrally buoyant case the simpler equations (\ref{eq:uG5})-(\ref{eq:omep5}).
In any case, the integrals of the preliminary velocity need to be computed. 
One of the benefits of using Lagrangian points also in the interior of the particles is that
these two integrals can be easily approximated using discrete sums
\begin{align}
\int_S \utilde \dd\boldsymbol{x} &\revision{=}{\,\approx} 
\sum_{\ell=1}^{N_L}\bm{\tilde{U}}(\bm{X}_\ell)\Delta V_{\ell},\label{eq:intutilde}\\
\int_S \bm{r}\times\utilde \mathrm{d}V &\revision{=}{\,\approx} 
\sum_{\ell=1}^{N_L}\bm{R}(\bm{X}_\ell)\times\bm{\tilde{U}}(\bm{X}_\ell)\Delta V_{\ell},\label{eq:intrvecutilde}
\end{align} 
\MGVN{where $\bm{R}(\bm{X}_\ell)=\bm{X}_\ell -\bm{x_p}$ is the $\ell$-th Lagrangian
point's position with respect to the particle's center of mass.}
\MGV{
If the distribution of Lagrangian points is sufficiently homogeneous and
uniform the discrete sums result in a second-order approximation to the integrals, which we have verified numerically. 
For a discussion about how to distribute points uniformly and associate them the corresponding
discrete volumes see
appendix B2 of \cite{moriche2021}. 
}
Introducing the above expressions into equations (\ref{eq:uG4})-(\ref{eq:omep4}), we obtain
\begin{align}
\uG^k&=
\left(1-\frac{\rhof}{\rhop}\right)\uG^{k-1}+
\frac{1}{V_c}\frac{\rhof}{\rhop}\sum_{\ell=1}^{N_L}\bm{\tilde{U}}(\bm{X}_\ell)\Delta V_{\ell}
+2\alpha_k\Delta t\left(1-\frac{\rhof}{\rhop}\right)\boldsymbol{g},\label{eq:uG6}\\
\omep^k&=\left(1-\frac{\rhof}{\rhop}\right)\omep^{k-1}
+\frac{1}{\tilde{I}_{p}}\frac{\rhof}{\rhop}
\sum_{\ell=1}^{N_L}
\left[
\bm{R}(\bm{X}_\ell)\times\bm{\tilde{U}}(\bm{X}_\ell)
\right]
\Delta V_{\ell}.
\label{eq:omep6}
\end{align}
In equation (\ref{eq:uG6}) the coefficient $2\alpha_k$ appears because the derivation of equation (\ref{eq:uG4}) was done for a generic interval with time step $\Delta t$, while the time step between the two Runge-Kutta stages $k-1$ and $k$ is  $2\alpha_k\Delta t$.
We compute now the new desired velocity by using the velocity field of the rigid body
\begin{equation}
\boldsymbol{U}^{(d)}(\boldsymbol{X}_\ell) = \uG^k + \omep^k\times \boldsymbol{R}(\boldsymbol{X}_\ell)
\quad\forall\ell.\label{eq:Udesired}
\end{equation}
Once the new desired velocity has been computed, the remaining steps of the original algorithm for the fluid phase are unmodified. Therefore, we have the following sequence of operations:
\begin{align}
&\boldsymbol{F}(\bm{X}_\ell)=
\frac{\boldsymbol{U}^{(d)}(\boldsymbol{X}_\ell)-\boldsymbol{\tilde{U}}(\boldsymbol{X}_\ell)}{\Delta t}\quad\forall\ell,\\
&f_\beta^k(\bm{x}_{ijk}^{\beta})=\sum_{\ell=1}^{N_L}F_\beta(\bm{X}_\ell)
\delta_h\left(\bm{x}_{ijk}^{\beta}-\bm{X}_\ell\right)\Delta V_\ell\quad\forall\,i,j,k;1\leq\beta\leq3, \label{eq:rk_fk}\\
&\nabla^2\bm{u}^*-\frac{\bm{u}^*}{\nu\alpha_k\Delta t}=-
\frac{1}{\nu\alpha_k}\left(
\frac{\utilde}{\Delta t}+\bm{f}^k
\right)
+\nabla^2\bm{u}^{k-1},\\
&\nabla^2\phi^k=\frac{\nabla\cdot\bm{u}^*}{2\alpha_k\Delta t},\\
&\bm{u}^k=\bm{u}^*-2\alpha_k\Delta t\nabla\phi^k,\\
&p^k=p^{k-1}+\phi^k-\alpha_k\Delta t\nu\nabla^2\phi^k,
\end{align}
where $\phi$ is the pseudo-pressure. Finally, we compute the new position of the centre of
mass of the particle
\begin{equation}
\bm{x_p}^k=\bm{x_p}^{k-1}+\alpha_k\Delta t\left(\uG^{k-1}+\uG^{k}
\right), 
\label{eq:rk_xG}
\end{equation}
and with this calculation the Runge-Kutta stage is completed. Note that
eq. (\ref{eq:rk_xG}) is independent of some of the previous steps and therefore 
can be computed \MGVN{at any point} after eq. (\ref{eq:rk_fk}). 
\MGV{As a conclusion, the overall algorithm comprises equations (\ref{eq:utilde}-\ref{eq:EuToLag},\ref{eq:uG6}-\ref{eq:rk_xG}) per Runge-Kutta stage.}

\MU{%
  While the scope of the present work is restricted to an isolated
  particle, let us note that the treatment of multi-particle
  configurations does not present any fundamental difficulty. More
  specifically, one needs to address the particle-particle (and
  particle-wall) interactions through some suitable contact model, such
  as e.g.\ a soft-sphere discrete element method 
  \citep{wachs:09,kempe:12b,kidanemariam:14a}. The force and torque arising
  through solid-solid contact are added to the right-hand-side of the
  Newton-Euler equations (\ref{eq:uG1}-\ref{eq:omep1}),
  and then these terms propagate along the derivation such that they
  enter the final algorithm in equations (\ref{eq:uG6}-\ref{eq:omep6}).
  A test of multiple interacting particles in the framework of the
  present formulation, however, is left for future work.
}
\section{Validation}

\subsection{Neutrally buoyant particle in 2D Couette flow} \label{sec:nb_couette}

\begin{figure}[ht]
  \centering
      \begin{minipage}{2.6ex}
\rotatebox{90}{\hspace{4ex}{$y_{eq}/H$}}
\end{minipage}
  \begin{minipage}{0.6\linewidth}
      \includegraphics[width=\linewidth,trim={4cm 8.5cm 4.2cm 9cm},clip]
      {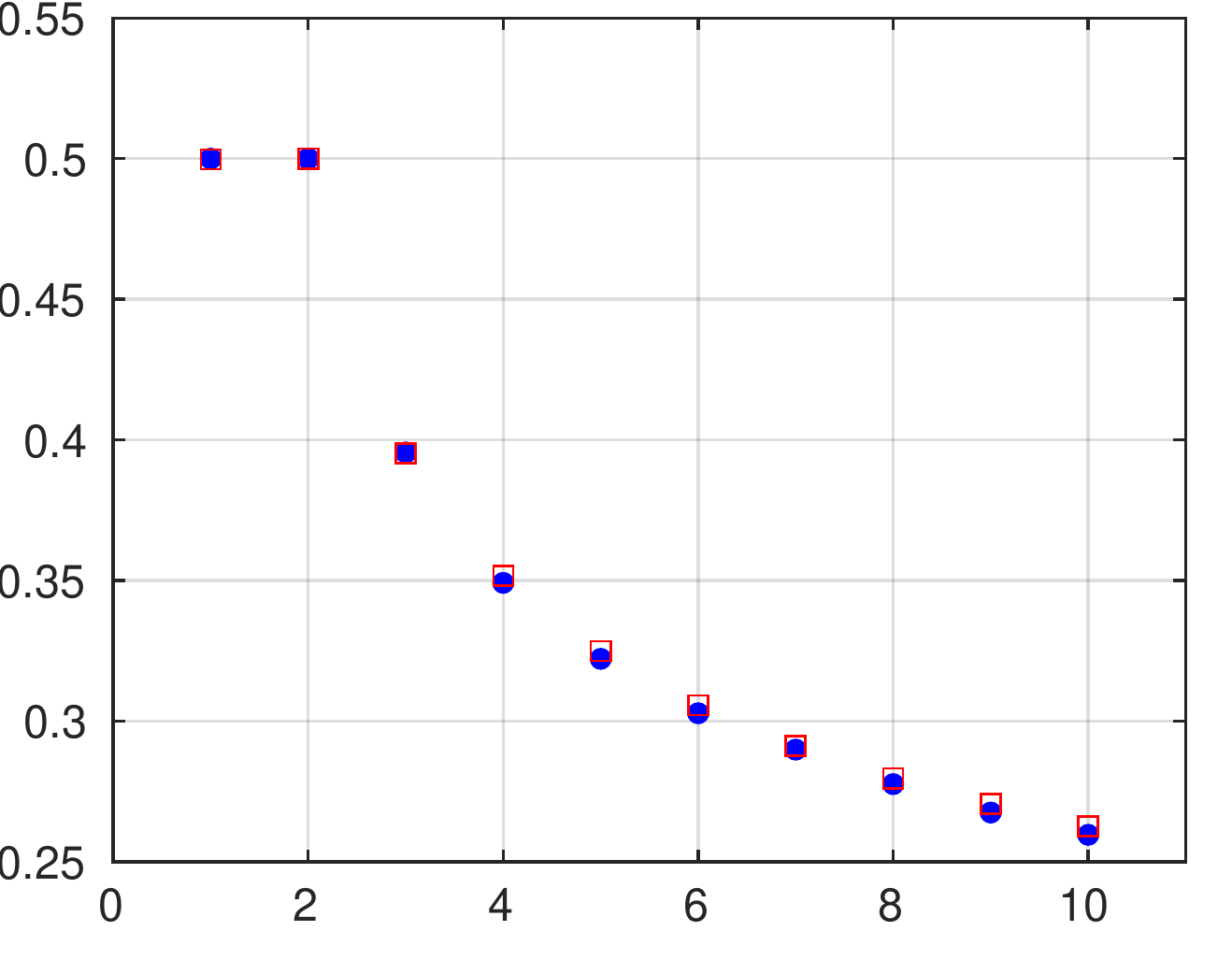}
            \centering{$Re_p$}
  \end{minipage}
\caption
{Equilibrium position, $y_{eq}/H$, as  a function of $Re_p$.
Red squares, present results for $D/\Delta x=75$. 
Blue circles, data digitized from reference \cite{pan2013}.
}
\label{fig:comp_yeq_pan2013}
\end{figure}

\begin{figure}[ht]
  \centering
    \begin{minipage}{2.6ex}
\rotatebox{90}{\hspace{2ex}{$y_p/H$}}
\end{minipage}
  \centering
  \begin{minipage}{0.45\linewidth}
      \raggedright{(\textit{a})}
      \includegraphics[width=\linewidth,trim={4cm 8.5cm 4.2cm 9cm},clip]
      {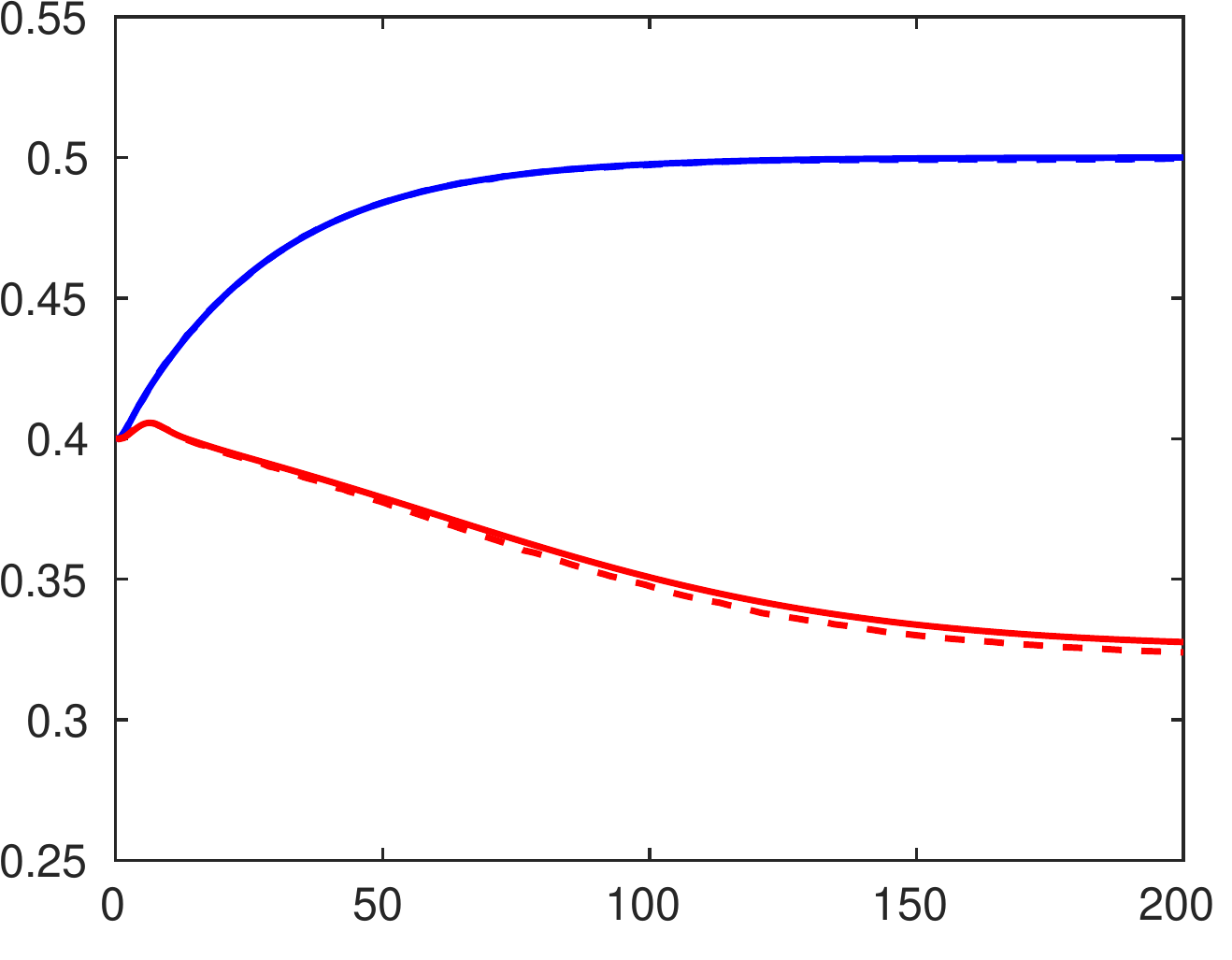}
      \centering{$t\Delta U/H$}
  \end{minipage}
  %
    \centering
    \begin{minipage}{2.6ex}
\rotatebox{90}{\hspace{2ex}{$\omega_P H/\Delta U$}}
\end{minipage}
  \centering
  \begin{minipage}{0.45\linewidth}
  		\raggedright{(\textit{b})}
  		\includegraphics[width=\linewidth,trim={4cm 8.5cm 4.2cm 9cm},clip]
      {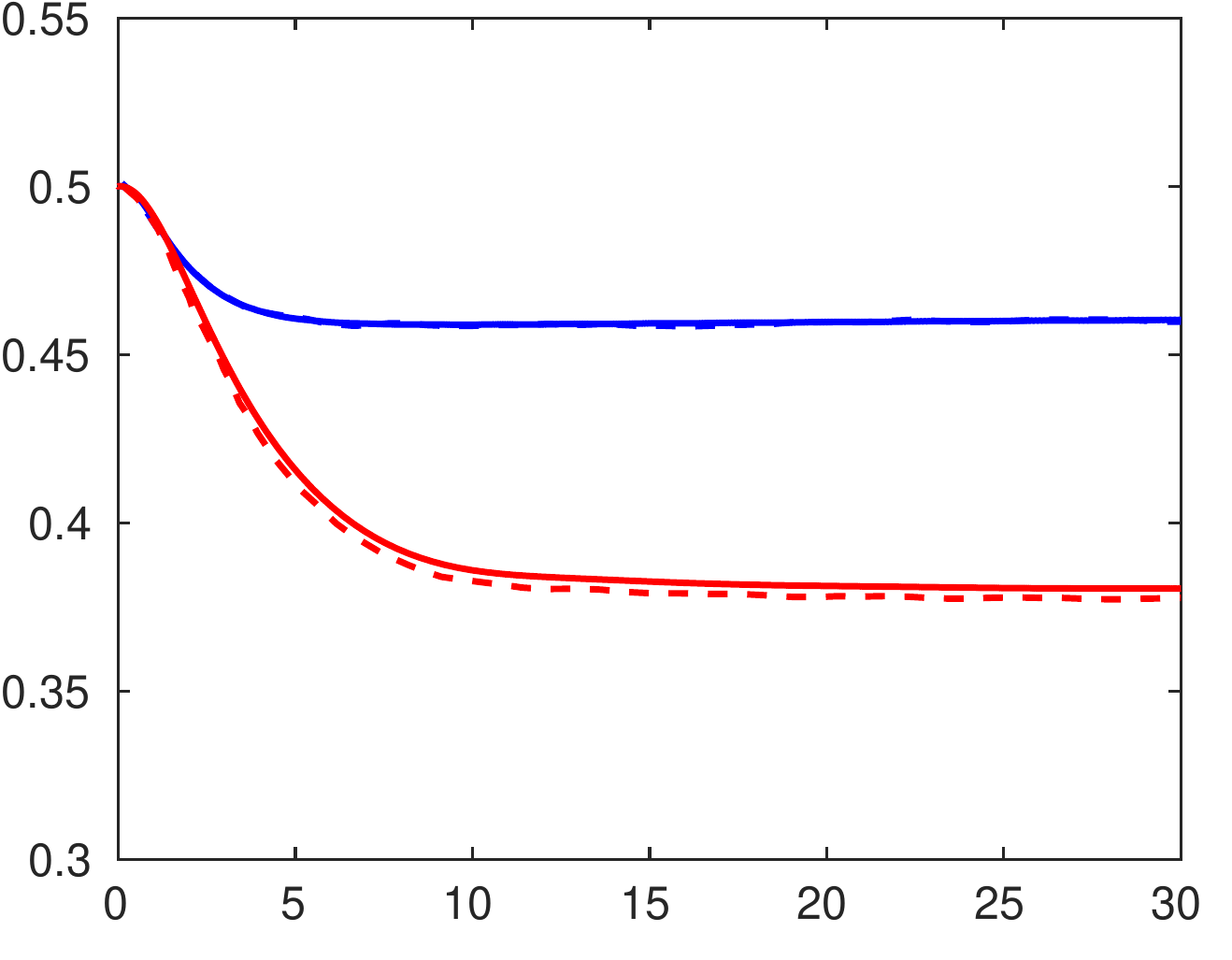}
        \centering{$t\Delta U/H$}
  \end{minipage}
\caption
{Time evolution of (\textit{a}) vertical position of particle center, $y_{G}/H$, and  (\textit{b}) particle
angular velocity, $\omega_p H /\Delta U$.
$Re_p=1$, Blue.
$Re_p=5$, Red.
Solid lines, present simulations with $D/\Delta x=50$.
Dashed lines, data digitized from Pan et al. \cite{pan2013}.
}
\label{fig:comp_pan_transient}
\end{figure}

As a first validation test case we have selected the lateral 
migration of a 
neutrally buoyant circular \MGV{particle} in 2D Couette flow  \cite{ding2000,pan2013,fox2020}.
When the neutrally buoyant particle is released from the centerline, the particle remains at that position.
However, if the particle is released off the centerline, it 
has been found that the particle has an additional equilibrium position
that depends on the particle Reynolds number \cite{pan2013,fox2020}.
In the present study we have \MGVN{chosen} the setup and data
of \citet{pan2013} as reference.
\revision{}{These authors employed a fictitious domain formulation with 
distributed Lagrange multipliers for the simulation of neutrally buoyant
circular and elliptic particles in 2D. 
Their computational methodology was based on finite element methods and an
operator splitting technique.}
The following notation and definitions are employed.
The distance between the walls is $H$ and their velocity difference is $\Delta U$. 
A neutrally buoyant particle, $\rhop/\rhof=1$, of radius $a=D/2=H/8$ 
is released at a distance to the wall $y_0=0.4H$,
\MGVN{
with initial linear and
angular velocities compatible with unperturbed Couette flow at $y_0$ 
(see the discussion at the end of  \S \ref{sec:coupling}).
}
We define the particle Reynolds number as 
$$
Re_p=\frac{\Delta U a^2}{H\nu},
$$
where $\nu$ is the kinematic viscosity. 
The length of the computational domain is $L_x=2H$. 
We impose no-slip conditions at the walls and periodicity along the streamwise direction. 
We employ a uniform grid with grid spacing $\Delta x$. 
We have performed several simulations varying the values of $Re_p$
in the range [1,10] for three grid spacings $D/\Delta x=25$, 50 and 75.
The corresponding time steps for each grid resolution are $\Delta t/(H/\Delta U)= 0.005$, 0.0025
and 0.00167, respectively.
For all the cases considered, after the particle is released it migrates laterally until
reaching an equilibrium position denoted by $y_{eq}$ with constant angular velocity $\omega_p$.
For $Re_p=1$ and 2, the particle settles at the centerline. 
For $Re_p\geq 3$, the equilibrium position is found off the centerline. 
The values of $y_{eq}$ for all cases are reported in Table \ref{tab:yeq} and a comparison
with the results of Pan et al. \cite{pan2013} is provided in Fig. \ref{fig:comp_yeq_pan2013},
showing a very good agreement.
The values of $\omega_p$ are reported in Table \ref{tab:omep}; this quantity was not available 
in the reference.
The time evolution of the vertical position of the particle center of mass, $y_p$, and the particle
angular velocity, $\omega_p$, are shown in Fig. \ref{fig:comp_pan_transient} for the two cases 
reported by Pan et al. \cite{pan2013}, $Re_p=1$ and 5. 
The agreement is also very good.

\begin{table}
\begin{center}
\begin{tabular}{l|cccccccccc}
$Re_p$          & 1    & 2      & 3      & 4      & 5      & 6      & 7      & 8      & 9      & 10 \\ \hline
$D/\Delta x=25$ & 0.5  & 0.4988 & 0.3962 & 0.3538 & 0.3272 & 0.3084 & 0.2946 & 0.2836 & 0.2746 & 0.4896 \\ 
$D/\Delta x=50$ & 0.5  & 0.5    & 0.3951 & 0.3516 & 0.3250 & 0.3058 & 0.2914 & 0.2801 & 0.2708 & 0.2632 \\ 
$D/\Delta x=75$ & 0.5  & 0.5    & 0.3952     & 0.3517     & 0.3249 & 0.3056     & 0.2912    & 0.2797     & 0.2705     & 0.2626 \\ \hline
\end{tabular}
\caption{Vertical equilibrium position $y_{eq}/H$ as a function of $Re_p$ and $D/\Delta x$. }
\label{tab:yeq}
\end{center}
\end{table}

\begin{table}
\begin{center}
\begin{tabular}{l|cccccccccc}
$Re_p$          & 1       & 2      & 3      & 4      & 5      & 6      & 7      & 8      & 9      & 10 \\ \hline
$D/\Delta x=25$ & 0.4612  & 0.4287 & 0.4094 & 0.3957 & 0.3845 & 0.3749 & 0.3666 & 0.3589 & 0.3522 & 0.3411 \\ 
$D/\Delta x=50$ & 0.4610  & 0.4281 & 0.4090 & 0.3955 & 0.3846 & 0.3754 & 0.3675 & 0.3605 & 0.3541 & 0.3483 \\ 
$D/\Delta x=75$ & 0.4608  & 0.4281 & 0.4088 & 0.3955 & 0.3845 & 0.3753     & 0.3675    & 0.3605     & 0.3542     & 0.3485 \\ \hline
\end{tabular}
\caption{Particle angular velocity $\omega_{p}H/\Delta U$ at equilibrium as a function of $Re_p$ and $D/\Delta x$. }
\label{tab:omep}
\end{center}
\end{table}

When comparing the data for the various grid resolutions reported in Tables \ref{tab:yeq} 
and \ref{tab:omep}, it is apparent that the steady-state quantities do not change much 
with the grid resolution.
In particular, the data  for $D/\Delta x=50$ and 75 are very close for all Reynolds numbers
considered. 
The lower resolution case, $D/\Delta x=25$, presents more deviations with respect to the higher
resolution cases. 
These deviations are not large, with the exception of the case $Re_p=10$ for which, in the 
low resolution case, the particle migrates to the centerline. 

When considering the transient behaviour, 
\MGVN{some} differences are observed among the various grid resolutions. 
When comparing the cases with $D/\Delta x=25$ and 50, \MGVN{small but visible} differences in the time evolution of $y_p$
appear for $Re_p\geq 3$.
When comparing the cases with $D/\Delta x=50$ and 75, 
differences in the time evolution of $y_p$
appear for $Re_p> 7$.
The trend is that the lower the resolution, the longer the time required for the particle
to settle at the equilibrium position.
These differences are not shown for the sake of brevity.

\subsection{Neutrally buoyant sphere released from rest in a free stream} \label{sec:nb_free_stream}

As a second test case, we have studied the release from rest of a neutrally buoyant sphere ($\rhop/\rhof=1$)
of diameter $D$ in a uniform free stream of speed $U$.
\revisions{This case has been previously studied by Tschisgale et al. \cite{tschisgale2017}.}
Reference data  for $Re_p=UD/\nu=20$ and 100 was obtained with 
a spectral element code as described in \ref{app:spectral}.
The simulations with the IBM proposed here are performed in a computational domain of size $L_x=30D$, $L_y=L_z=5.34D$. 
The lateral dimensions coincide with those of the study of \cite{uhlmann2014}.
At the inlet boundary we impose a Dirichlet boundary condition, $(u,v,w)=(U,0,0)$, at the outflow boundary a convective boundary condition and at the lateral boundaries free-slip boundary
conditions. 
We have performed simulations at $Re_p=20$ and 100, with grid resolutions $D/\Delta x=18$, 24 and 36. The time step has been adapted so that the $CFL$ number is approximately 0.1. 

The simulations are performed in two phases. 
In the first phase, the flow is computed with the particle fixed at a distance to the inlet plane $x_0=5D$. 
Once the 
\MGVN{flow has reached a steady state (i.e the drag force becomes time independent)}, we release the particle which is then accelerated
\MGVN{until it ultimately reaches an equilibrium state with $u_p=U$.}
Table \ref{tab:cd_sph_fixed_norot} provides the drag coefficient of the particle before it is 
released, for both values of $Re_p$. 
The variation with the grid resolution is small in all cases. For $Re_p=20$ ($Re_p=100$),
$C_D$ changes by 0.4\% (0.2\%) between $D/\Delta x=18$ and $D/\Delta x=36$.
The differences observed when comparing the IBM data and the
spectral element data are somewhat larger.
For $Re_p=20$ ($Re_p=100$),
$C_D$ differs by 1.8\% (2.2\%) between the IBM case with $D/\Delta x=36$ and 
the spectral element data. 
\revision{
Note that these variations are still reasonably small and may partially be attributed
to the larger  computational domain of the reference simulation, namely the larger distance to the inflow boundary.}
{Note that these variations are still reasonably small, and that
similar differences have already been observed in previous studies
where results from boundary-conforming spectral/spectral-element
simulations in cylindrical domains were compared to those from
immersed-boundary-type non-conforming methods on Cartesian grids 
\citep{uhlmann2014,rettinger:17,moriche2021}.  
The remaining differences can be attributed to the larger
computational domain of the reference simulation (featuring a larger
distance to the inflow boundary), to the different shape of the
domain, and to the different lateral boundary conditions.}

\begin{table}[ht]
\begin{center}
\begin{tabular}{l|cccc}
          & Reference    & $D/\Delta x=18$    & $D/\Delta x=24$  & $D/\Delta x=36$ \\ \hline
$Re=20$   &   2.7753     &     2.8152         &     2.8195       &   2.8261         \\ 
$Re=100$  &   1.0962     &     1.1179         &     1.1185       &   1.1201         \\ \hline
\end{tabular}
\caption{Drag coefficient, $C_D$, of a fixed sphere in a uniform flow as a function of Reynolds number, $Re$,  and grid resolution, $D/\Delta x$. Reference data corresponds to the computations with the spectral element code.}
\label{tab:cd_sph_fixed_norot}
\end{center}
\end{table}

\begin{figure}[ht]
  \centering
    \begin{minipage}{2.6ex}
\rotatebox{90}{\hspace{2ex}{$u_p/U$}}
\end{minipage}
  \centering
  \begin{minipage}{0.46\linewidth}
      \raggedright{(\textit{a})}
      \includegraphics[width=\linewidth,trim={3.4cm 8.7cm 4cm 9cm},clip]
      {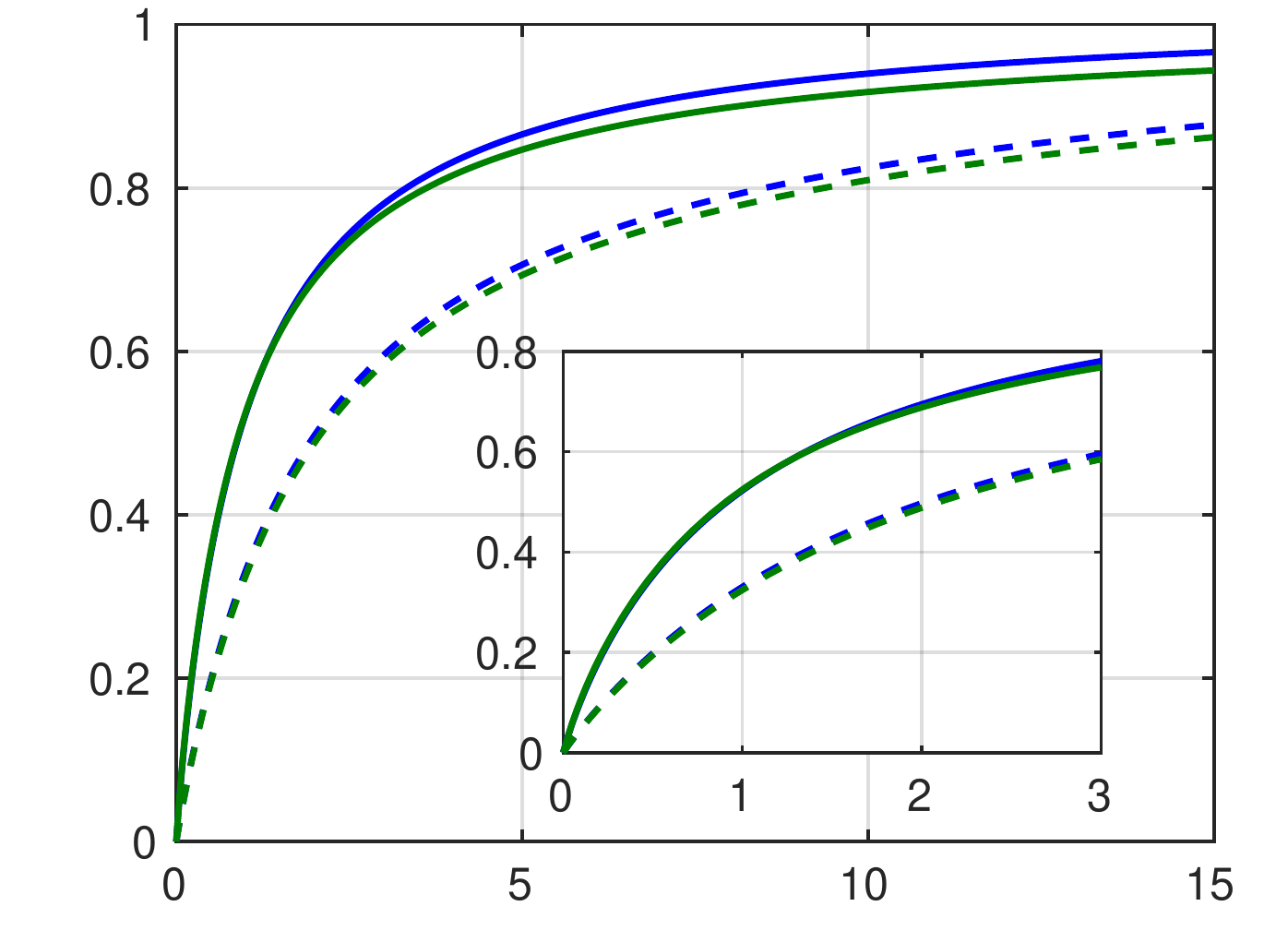}
              \centering{$t\, U/D$}
  \end{minipage}
  %
    \centering
    \begin{minipage}{2.6ex}
\rotatebox{90}{\hspace{2ex}{$\epsilon_{u}$}}
\end{minipage}
  \centering
  \begin{minipage}{0.46\linewidth}
  		\raggedright{(\textit{b})}
  		\includegraphics[width=\linewidth,trim={3.4cm 8.7cm 4cm 9cm},clip]
      {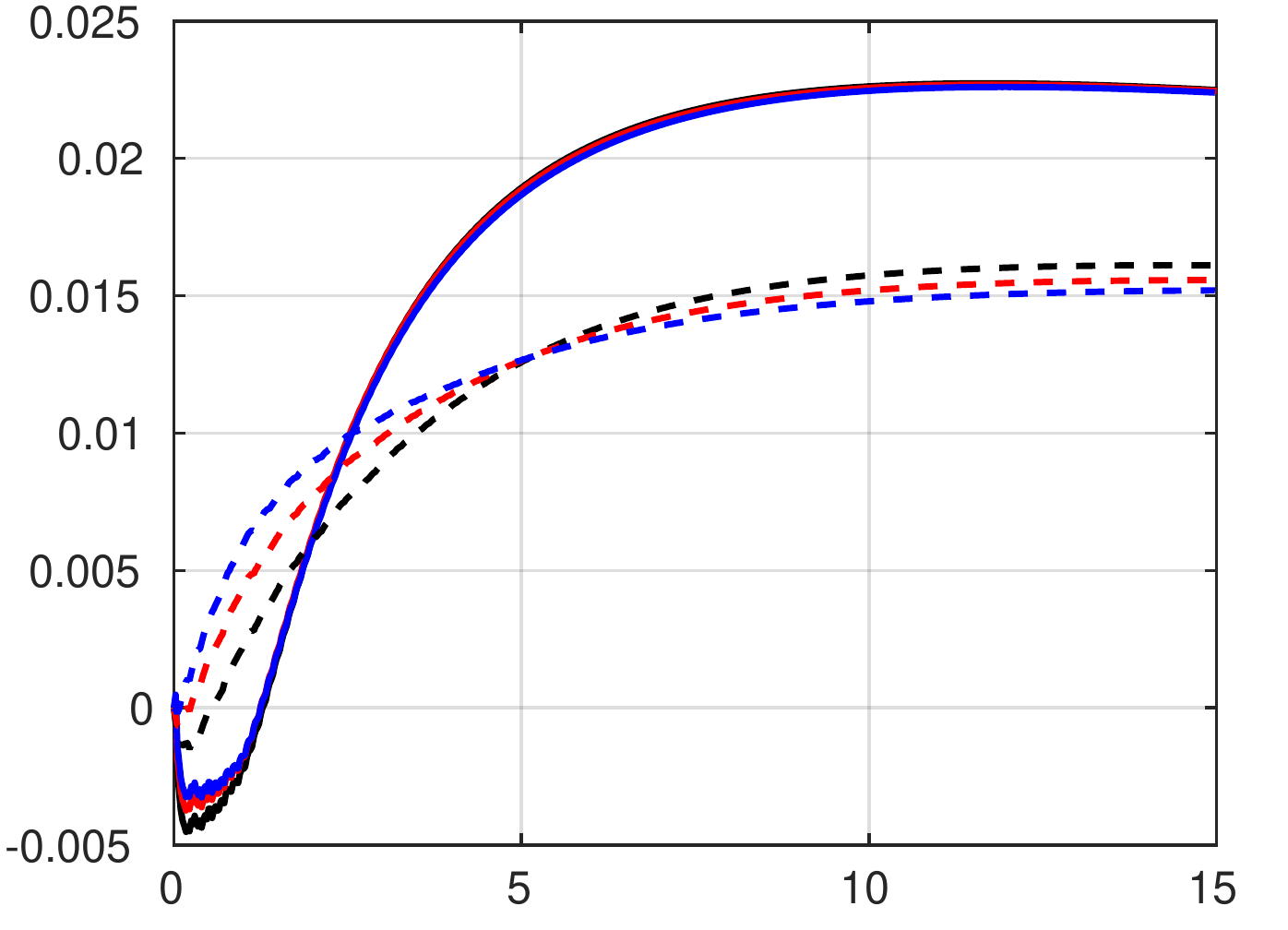}
              \centering{$t\, U/D$}      
  \end{minipage}
\caption
{Time evolution of (\textit{a}) particle streamwise velocity component, $u_{p}/U$, and  (\textit{b}) error with 
respect to reference data, $\epsilon_{u}$.
Green lines, reference data. 
$Re_p=20$, solid lines.
$Re_p=100$, dashed lines.
$D/\Delta x=18$, Black.
$D/\Delta x=24$, Red.
$D/\Delta x=36$, Blue.
Inset in (\textit{a}) highlights the initial phase.
}
\label{fig:comp1_dusek}
\end{figure}

Fig. \ref{fig:comp1_dusek}$a$ shows the time evolution of the  particle streamwise
velocity component, $u_p$, for both Reynolds numbers. 
The lower the Reynolds number, the faster the velocity of the particle approaches
the free-stream velocity. 
This effect is well captured by the simulations with the IBM method. 
The agreement with the reference data is good although some differences are visible in the
figure.
In order to quantify these differences, Figure \ref{fig:comp1_dusek}$b$,
shows the error of the streamwise component of the particle's velocity
with respect to the reference data,
$\epsilon_{u}$, 
as a function of time.
\MU{%
  Note that the normalized error for any quantity $\phi$ here and in
  the following is defined as  
  \begin{equation}\label{equ-error-definition}
    \epsilon_{\phi}(t)=\left(\phi(t)-\phi^{ref}(t)\right)/\phi_{norm}
    \,,
  \end{equation}
  where $\phi^{ref}$ is the reference value and $\phi_{norm}$ the
  scale used for normalization. In measuring the error of the
  streamwise particle velocity ($\phi=u_p$) we use $\phi_{norm}=U$. 
}

The data
\MU{in Figure~\ref{fig:comp1_dusek}$b$} 
shows that the IBM simulations are \MGV{generally} in better agreement with the reference simulations for $Re_p=100$ than for $Re_p=20$. 
\MGVN{
For long times, $t\,U/D>5$ the  error presents a plateau with a value of about
2.3\% (1.5\%)  for $Re_p=20$ ($Re_p=100$).
%
%
The effect of the grid resolution is barely visible for $Re_p=20$ with the three solid lines collapsing in Figure  \ref{fig:comp1_dusek}$b$, except for very short times,  $t\,U/D<1$.
In the case of  $Re_p=100$, some differences are visible, with the temporal evolution of the error showing a cross-over at about $t\,U/D\approx5$.
The finest grid with $D/\Delta x=36$ presents the lowest error of the three 
in the plateau region, although the differences are small.
This indicates that, for this configuration, a grid resolution of
$D/\Delta x=18$ provides reasonable accuracy for simulations at these Reynolds numbers.
}

\begin{figure}[ht]
  \centering
    \begin{minipage}{2.6ex}
\rotatebox{90}{\hspace{2ex}{$u_p/U$}}
\end{minipage}
  \centering
  \begin{minipage}{0.46\linewidth}
      \raggedright{(\textit{a})}
      \includegraphics[width=\linewidth,trim={3.4cm 8.7cm 4cm 9cm},clip]
      {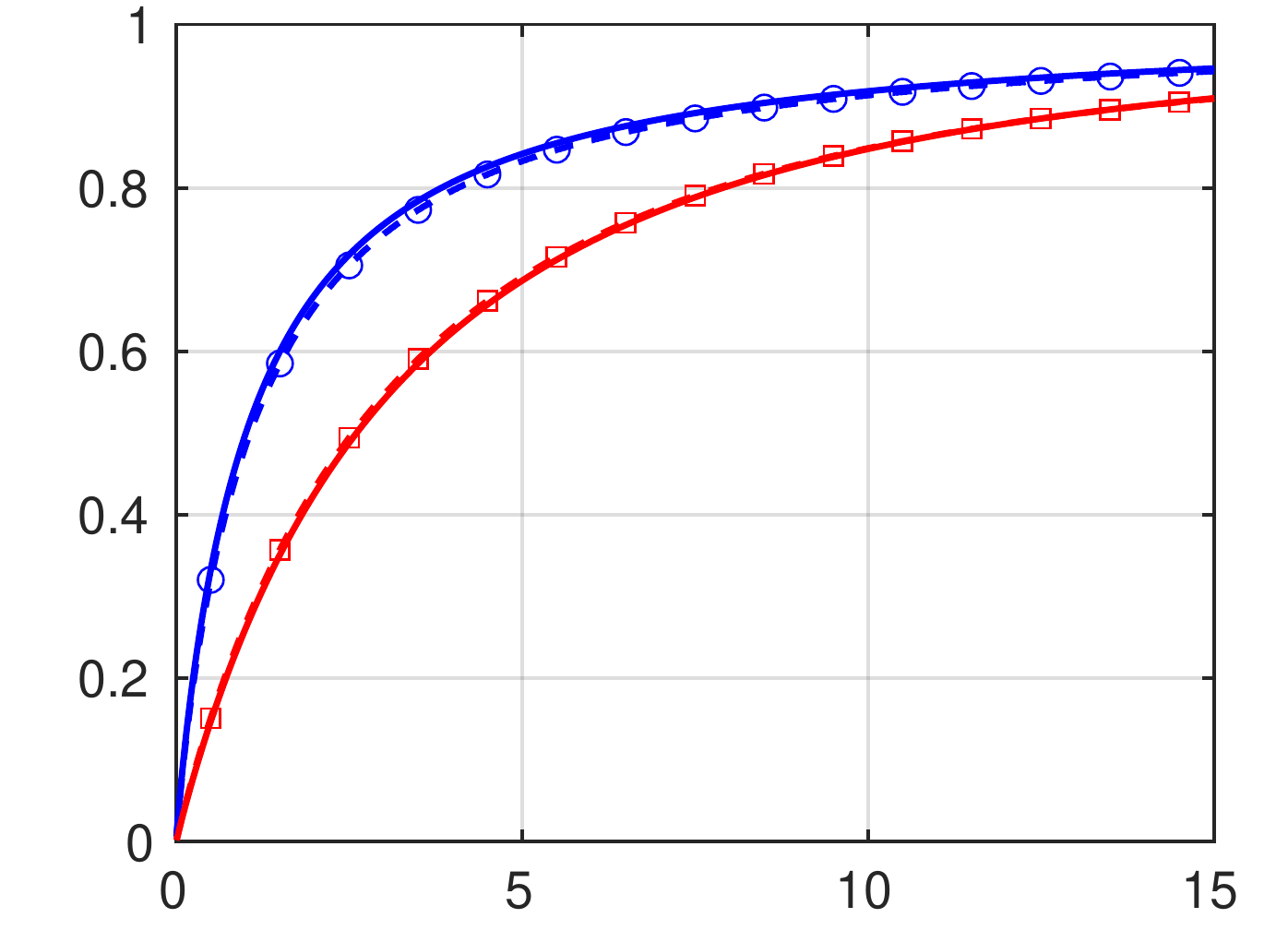}
              \centering{$t\, U/D$}
  \end{minipage}
  %
    \centering
    \begin{minipage}{2.6ex}
\rotatebox{90}{\hspace{2ex}{$u_p/U$}}
\end{minipage}
  \centering
  \begin{minipage}{0.46\linewidth}
  		\raggedright{(\textit{b})}
  		\includegraphics[width=\linewidth,trim={3.4cm 8.7cm 4cm 9cm},clip]
      {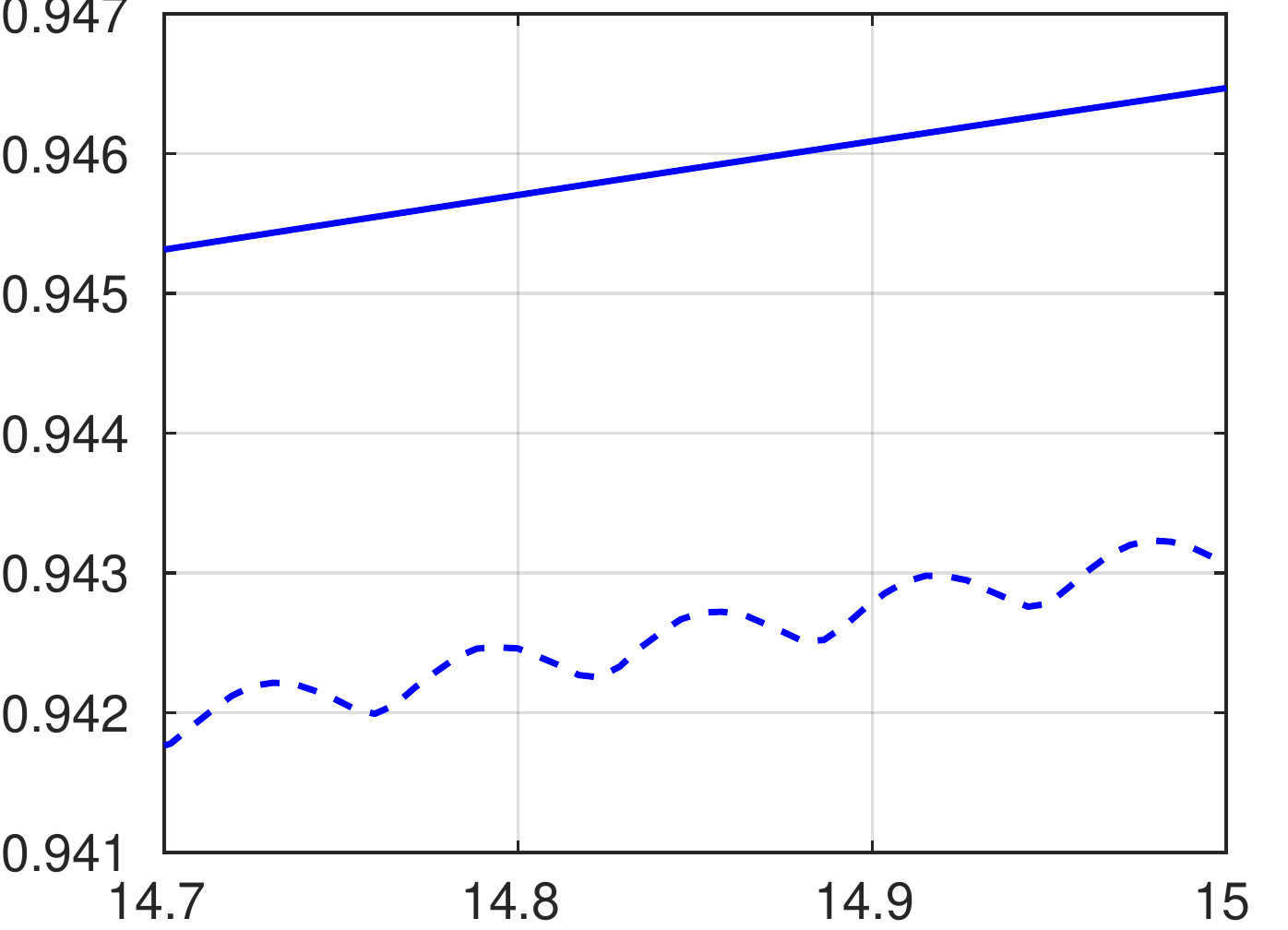}
              \centering{$t\, U/D$}      
  \end{minipage}
\caption
{\revision{}{(\textit{a}) Time evolution of particle streamwise velocity component, $u_{p}/U$.
Dashed lines with symbols, reference data from \citet{tschisgale2017}.
Solid lines , present results.
Blue, $\rho_p/\rho_f=1.05$.
Red, $\rho_p/\rho_f=5$.
(\textit{b}) Zoom of  
(\textit{a}).
}
}
\label{fig:comp_tschisgale}
\end{figure}

\revision{}{A very similar configuration has been studied 
by other authors \citep{kempe2012,schwarz2015,tschisgale2017}.
Here we have reproduced their setup to provide a direct assessment
of the present methodology. 
The size of the computational domain is $L_x=30D$, $L_y=L_z=15D$.
The center of the particle is initially located at 10$D$ from the inflow plane
and centered with respect to the lateral planes.
The number of grid points is $512\times 256\times 256$ which corresponds 
to a grid resolution of $D/\Delta x\approx 17$.
The Reynolds number is $Re_p=20$ and the time step has been adapted so that the 
$CFL$ number is approximately 0.1.
With respect to the simulations discussed above, an additional difference is that
Dirichlet boundary conditions are applied at the lateral boundaries,
where the velocity of the fluid is set equal to the free stream velocity, $U$.
Two simulations have been performed for the density ratios
$\rhop/\rhof=1.05$ and $5$.
Fig. \ref{fig:comp_tschisgale}$a$ shows the time evolution of the  particle streamwise
velocity component, $u_p$, for both density ratios.
The present results are in good agreement 
with the data from \citet{tschisgale2017}.
Fig. \ref{fig:comp_tschisgale}$b$ shows
a zoom of panel $(a)$ towards the end
of the simulated interval. 
This figure illustrates that the present
method is free of spurious oscillations.
}

\subsection{Sphere with initial rotation released in a free stream} \label{sec:rot}

The test case considered in the previous section is rather simple 
with unidirectional motion and not involving the rotation of the particle. 
Therefore, for the final test case, we have tried to avoid those two limitations
by defining a test case not reported before. 
First, while the particle is kept at a fixed location \MGV{with a uniform inflow velocity}, 
a \MGV{constant angular particle} velocity is imposed so that both lift and drag develop. 
Once the lift and drag \MGV{forces have} converged, we release the particle
\MGV{(at an instant $t=t_1$ that we arbitrarily set to zero, i.e. $t_1=0$)}.
In the subsequent evolution, the particle is accelerated in the streamwise direction while, due 
to the initial lift, a lateral motion is generated. 
As a consequence, the particle experiences both rotation and lateral motion, which are both 
damped as time evolves.

We have performed simulations with $Re_p=100$ and with an initial angular velocity $\omega_{pz}(t\leq t_1)=-\Omega$ with $\Omega D/U =1$.
Reference data was again obtained with the spectral element code described in \ref{app:spectral}.
The computational domain and boundary conditions at inflow and outflow boundaries \MGV{used in the IBM simulations} are the same
as in the previous section. 
However, due to the lateral motion it was not possible to use 
free-slip conditions at the lateral boundaries
since the particle may approach one of the boundaries \MGVN{after it has been released}. 
As a consequence, periodic conditions have been employed at the lateral boundaries.
The grid resolution has been varied as in the previous section, with values $D/\Delta x=18, $ 24 and 36, and the time step has been adjusted so that $CFL\approx 0.1$.
In this section, we consider neutrally buoyant particles ($\rhop/\rhof=1$) and also
particles with other density ratios to illustrate the capabilities of the proposed method.
Note that gravity is set to zero in all cases, so that the effect of the density ratio only
affects the problem via the inertia term.

\begin{figure}[ht]
  \centering
    \begin{minipage}{2.6ex}
\rotatebox{90}{\hspace{2ex}{\large $y/D$}}
\end{minipage}
  \begin{minipage}{0.5\linewidth}
      \raggedright{(\textit{a})}
      \includegraphics[width=\linewidth,trim={2cm 0.5cm 2cm 1.2cm},clip]
      {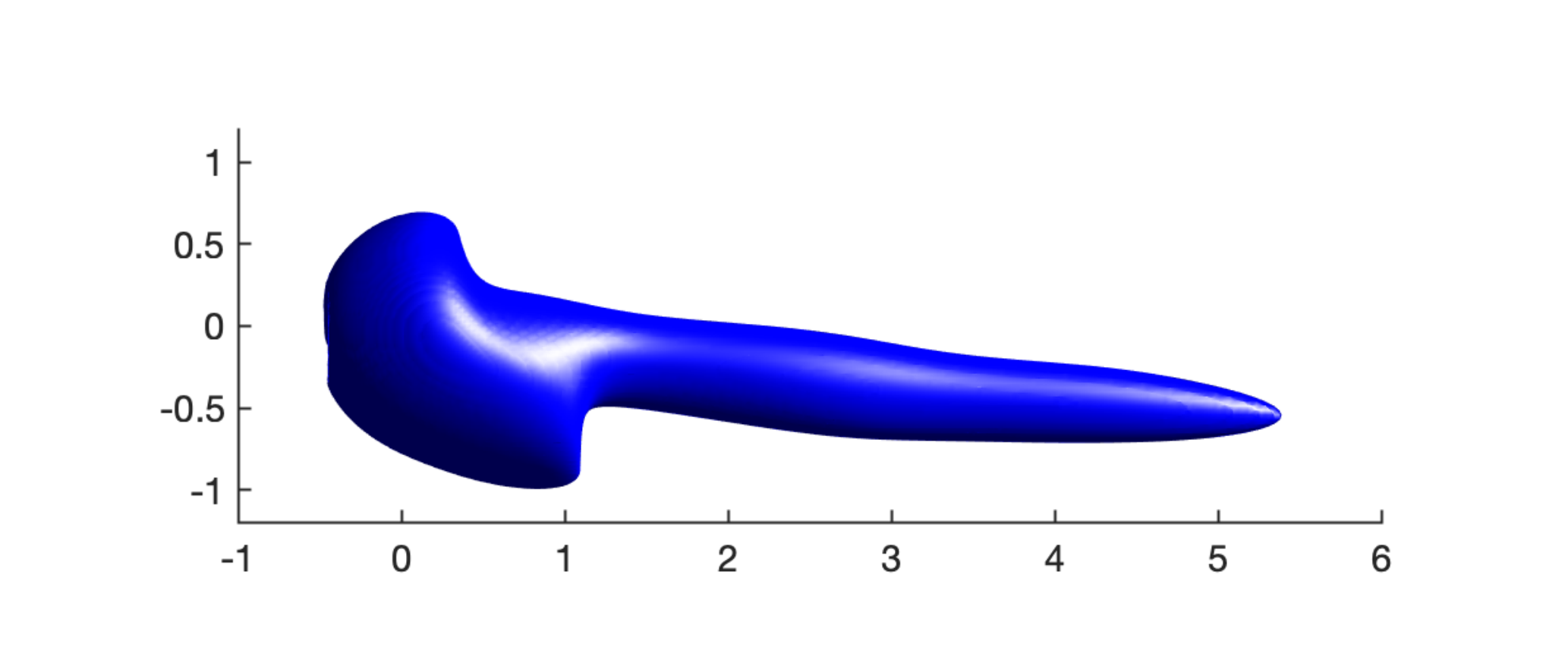}
  \end{minipage}
  \\
    \begin{minipage}{2.6ex}
\rotatebox{90}{\hspace{2ex}{\large $z/D$}}
\end{minipage}
  \begin{minipage}{0.5\linewidth}
      \raggedright{(\textit{b})}
      \includegraphics[width=\linewidth,trim={2cm 0.5cm 2cm 1.2cm},clip]
      {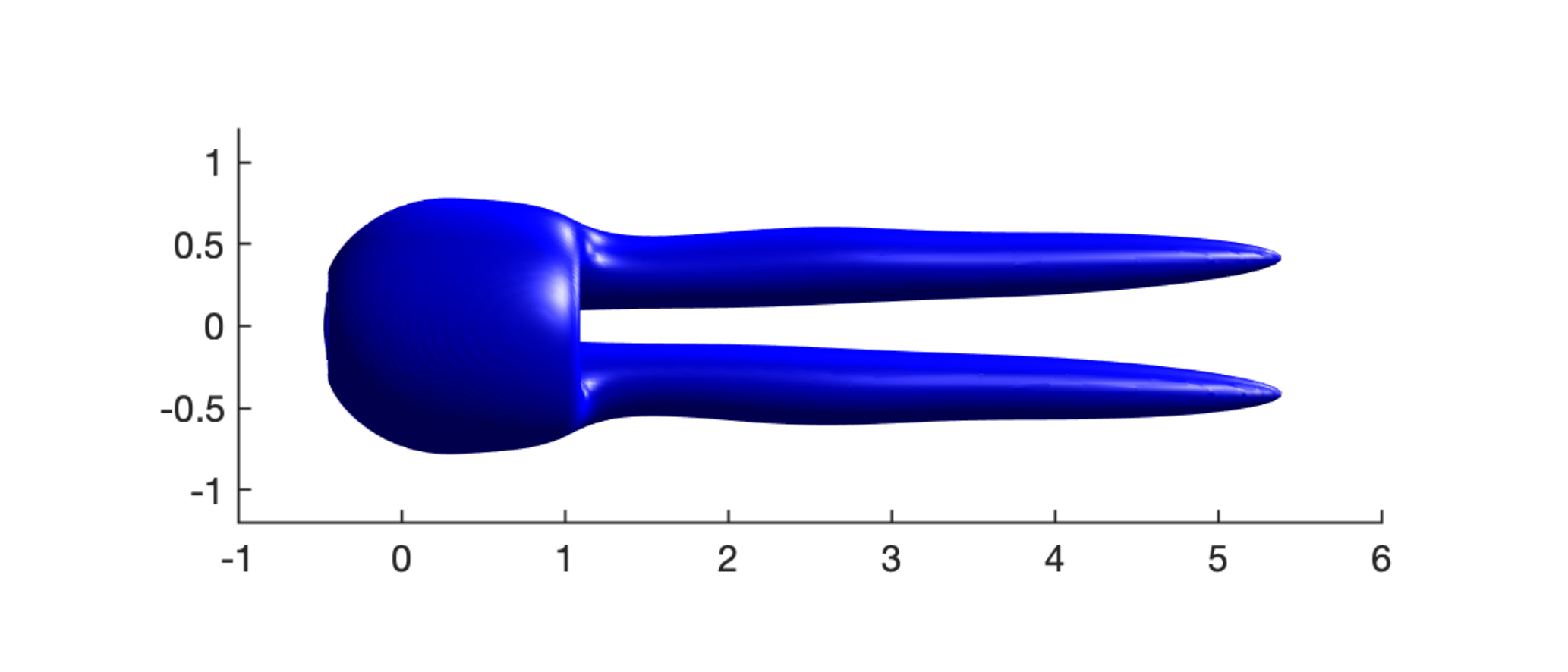}
  \end{minipage}
    \begin{minipage}{0.5\linewidth}
      \centering{{\large $x/D$}}
  \end{minipage}
\caption
{Flow visualization of wake structure before the rotating particle is released. Two side views
of an iso-surface of the second invariant of the velocity gradient tensor, \MGV{$Q=0.05 U^2/D^2$.}
\MGV{Data correspond to the IBM simulation with $D/\Delta x = 36$. }}
\label{fig:isoQ_ic}
\end{figure}

First, let us characterize the initial state before the particle is released.
Table \ref{tab:cd_sph_fixed_rot} provides the drag and lift coefficients.
The effect of grid resolution is once again small: $C_D$ ($C_L$) varies \MGV{by approximately}
0.1\% (0.5\%) when comparing the simulations with $D/\Delta x=18$ and 36.
When comparing with the reference data, we observe that the lift coefficient 
is closer to the reference data (0.5\% difference) than the drag coefficient (2.7\% difference).
Note that the difference in drag coefficient is of the same order of magnitude as the one
observed for the case without rotation of the particle in \S \ref{sec:nb_free_stream}.
This might indicate that the drag coefficient may be more influenced by the 
differences in shape of the computational domain and lateral boundary conditions
than the lift coefficient, the latter being \MGV{caused by} the rotation 
of the particle and therefore having a more local origin. 
The wake structure of the rotating particle \MGVN{(while it is still held fixed)} is visualized from two angles in Figure \ref{fig:isoQ_ic} 
with an iso-surface of the second invariant of the velocity gradient tensor, $Q$.
Note that the rotation axis is the $z-$axis so that the lift \MGV{force points in} the $y-$direction.
Consequently, when the flow is visualized from the $z-$axis, Figure \ref{fig:isoQ_ic}$a$,
an asymmetry with respect to the $xz-$plane is observed. 
When the flow is visualized from the $y-$axis, Figure \ref{fig:isoQ_ic}$b$,
the double threaded structure of the wake is revealed.

\begin{table}
\begin{center}
\begin{tabular}{l|cccc}
          & Reference    & $D/\Delta x=18$    & $D/\Delta x=24$  & $D/\Delta x=36$ \\ \hline
$C_D$   &   1.2217     &     1.2539        &     1.2540       &   1.2550         \\ 
$C_L$  &    0.4972     &     0.4973        &     0.4990       &   0.5002         \\ \hline
\end{tabular}
\caption{Drag and lift coefficients of a fixed, rotating sphere in a uniform flow  as a function of grid resolution, $D/\Delta x$. Reference data corresponds to the computations with the spectral element code.}
\label{tab:cd_sph_fixed_rot}
\end{center}
\end{table}

\begin{figure}[ht]
  \centering
  \begin{minipage}{2.6ex}
\rotatebox{90}{\hspace{2ex}{\Large $y/D$}}
\end{minipage}
  \begin{minipage}{0.9\linewidth}
      \raggedright{(\textit{a})}
      \includegraphics[width=\linewidth,trim={2cm 1.15cm 2cm 1.2cm},clip]
      {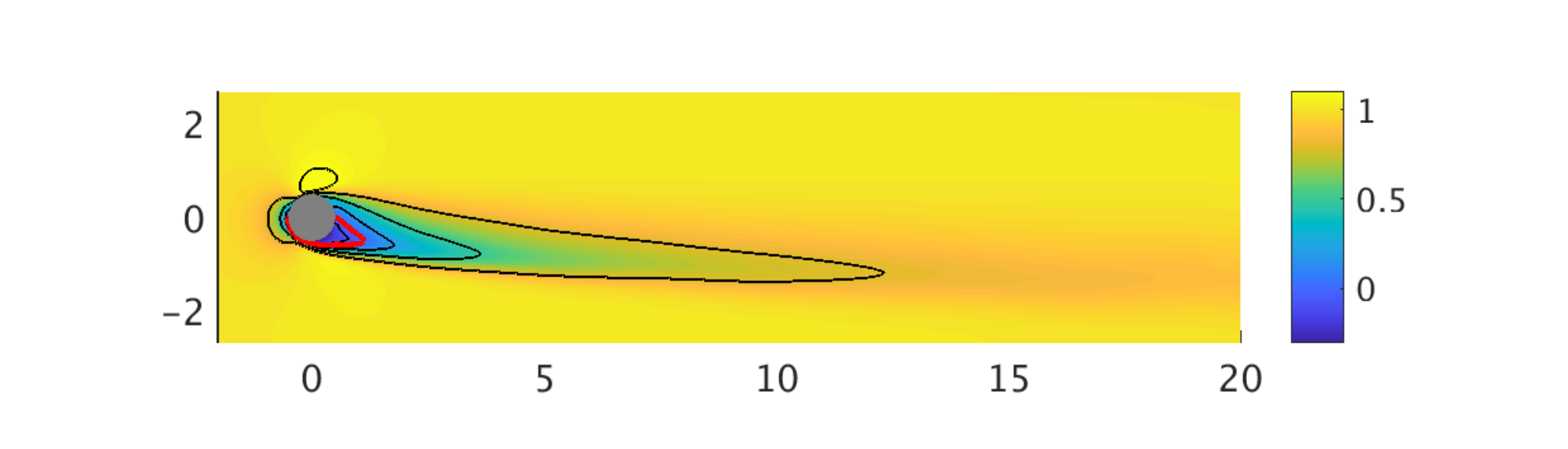}
  \end{minipage}
  \\
    \begin{minipage}{2.6ex}
\rotatebox{90}{\hspace{2ex}{\Large $y/D$}}
\end{minipage}
  \begin{minipage}{0.9\linewidth}
      \raggedright{(\textit{b})}
      \includegraphics[width=\linewidth,trim={2cm 1.15cm 2cm 1.2cm},clip]
      {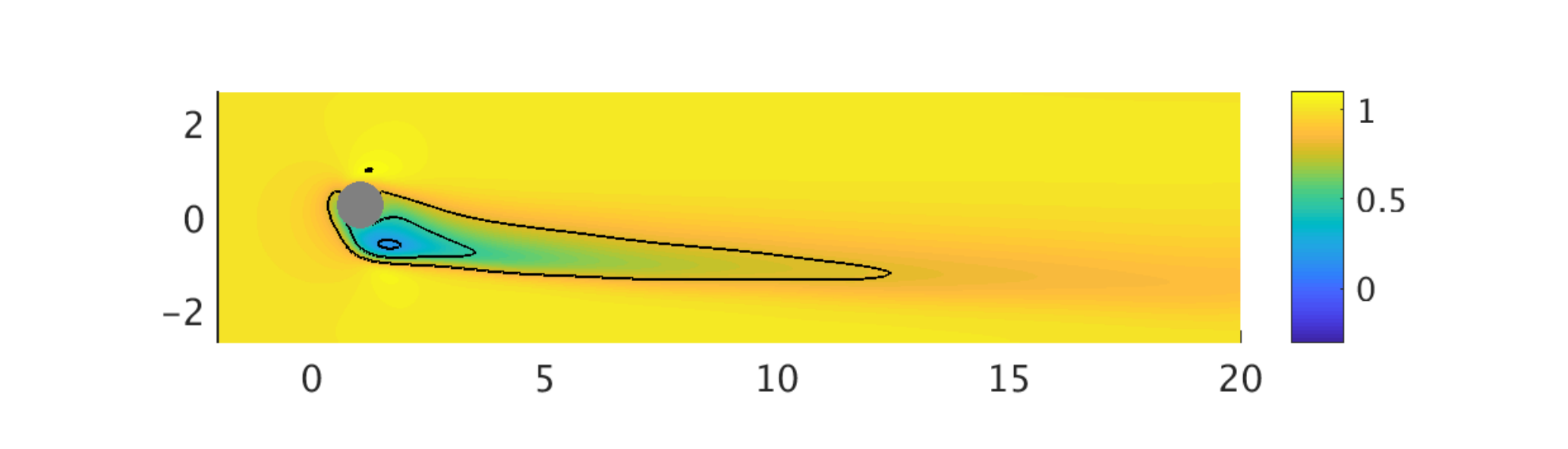}
  \end{minipage}
 \\
    \begin{minipage}{2.6ex}
\rotatebox{90}{\hspace{2ex}{\Large $y/D$}}
\end{minipage}
  \begin{minipage}{0.9\linewidth}
      \raggedright{(\textit{c})}
      \includegraphics[width=\linewidth,trim={2cm 1.15cm 2cm 1.2cm},clip]
      {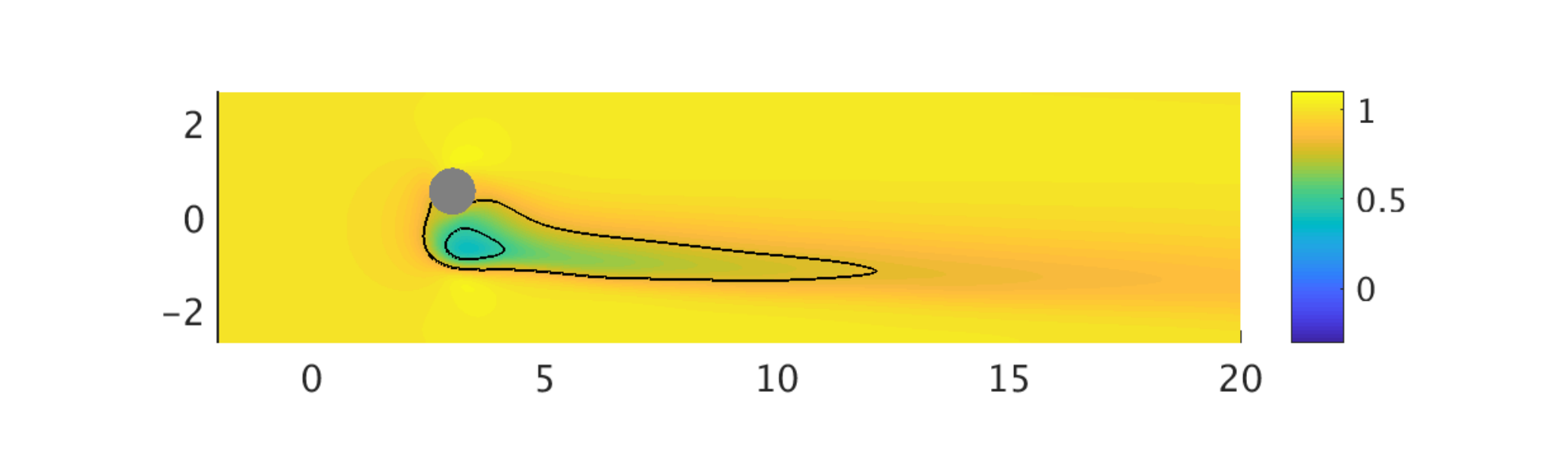}
  \end{minipage}
 \\
    \begin{minipage}{2.6ex}
\rotatebox{90}{\hspace{2ex}{\Large $y/D$}}
\end{minipage}
  \begin{minipage}{0.9\linewidth}
      \raggedright{(\textit{d})}
      \includegraphics[width=\linewidth,trim={2cm 1.15cm 2cm 1.2cm},clip]
      {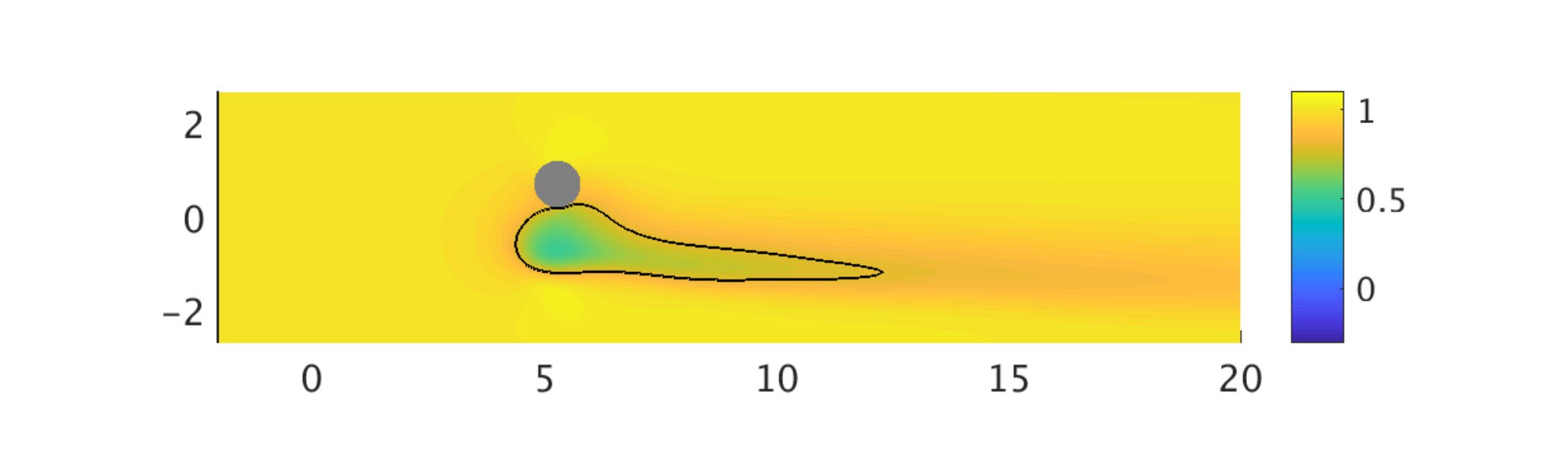}
  \end{minipage}
 \\
    \begin{minipage}{2.6ex}
\rotatebox{90}{\hspace{2ex}{\Large $y/D$}}
\end{minipage}
  \begin{minipage}{0.9\linewidth}
      \raggedright{(\textit{e})}
      \includegraphics[width=\linewidth,trim={2cm 0.5cm 2cm 1.2cm},clip]
      {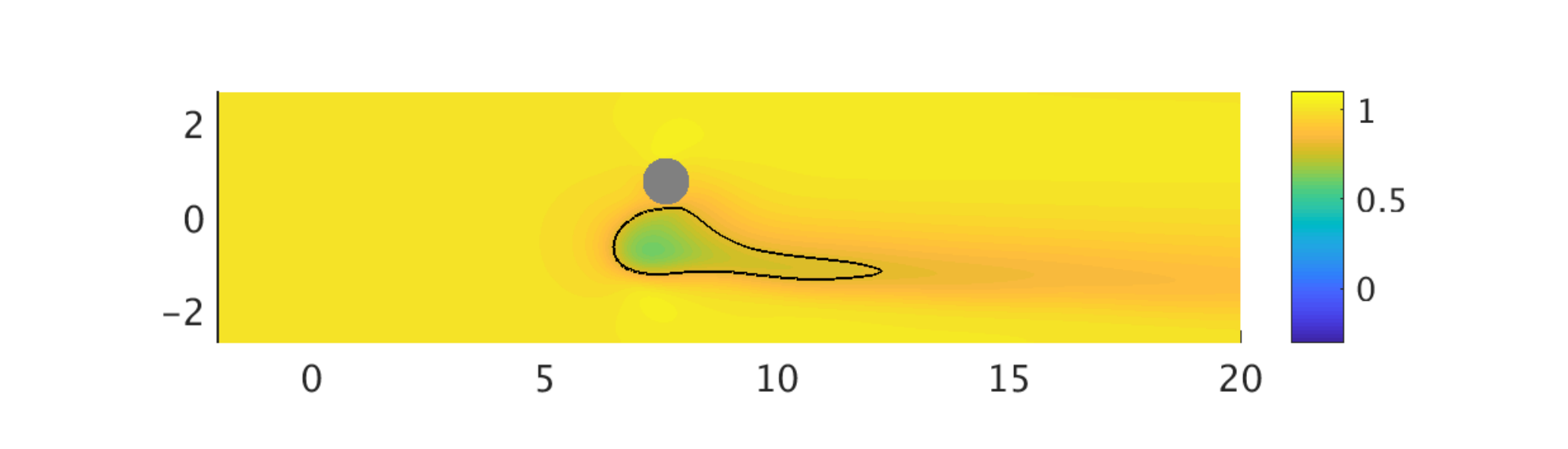}
  \end{minipage}
    \begin{minipage}{0.9\linewidth}
      \centering{{\Large $x/D$}}
  \end{minipage}

\caption
{Contours of streamwise velocity of the fluid phase, $u$,  in the plane $z=0$ \MGV{for the neutrally buoyant particle which is released
after initially rotating at constant angular velocity.
Data correspond to the IBM simulation with $D/\Delta x = 36$. }
From top to bottom, the time instants are
$t\, U/D=0$, 2.5, 5, 7.5 and 10 \MGV{after particle release}. The red line in panel ($a$) corresponds to $u/U=0$.
The black lines correspond to $u/U=[-0.1, 0.2, 0.5, 0.8, 1.1]$.
}
\label{fig:u2d_evol}
\end{figure}

\MGV{Let us first consider the case of} a neutrally buoyant particle.
Figure \ref{fig:u2d_evol} shows snapshots of the streamwise velocity of the fluid phase in the plane
$z=0$ at several instants. 
We restrict the analysis to the $z=0$ plane since the particle does not move laterally along
the $z-$axis as time evolves, being $w_p/U=0$ for all times.
Before the particle is released (panel $a$) the fluid velocity in this plane
is in the range $u/U\in[-0.3,1.2]$ with the smallest values concentrated directly behind the sphere. 
As time evolves \MGV{subsequent to the particle's release,} the fluid velocity approaches the free stream velocity $U$. 
As an example, for $t\,U/D=10$ (panel $e$) the fluid velocity is in the range $u/U\in[0.6,1.05]$.
Once the particle is released, it starts moving downstream, \MGV{while also migrating} laterally towards
positive values of the $y$ coordinate.
\MGV{Thus, the particle becomes detached from its initial wake. }
%
As time evolves the particle separates \MGVN{further} from the wake and gradually \MGVN{passes}
the region with low fluid velocities, as clearly seen in panels ($c$) to ($e$). 

\begin{figure}[ht]
  \centering
    \begin{minipage}{2.6ex}
\rotatebox{90}{\hspace{2ex}{$u_p/U$}}
\end{minipage}
  \centering
  \begin{minipage}{0.46\linewidth}
      \raggedright{(\textit{a})}
      \includegraphics[width=\linewidth,trim={3.4cm 8.7cm 4cm 8.5cm},clip]
      {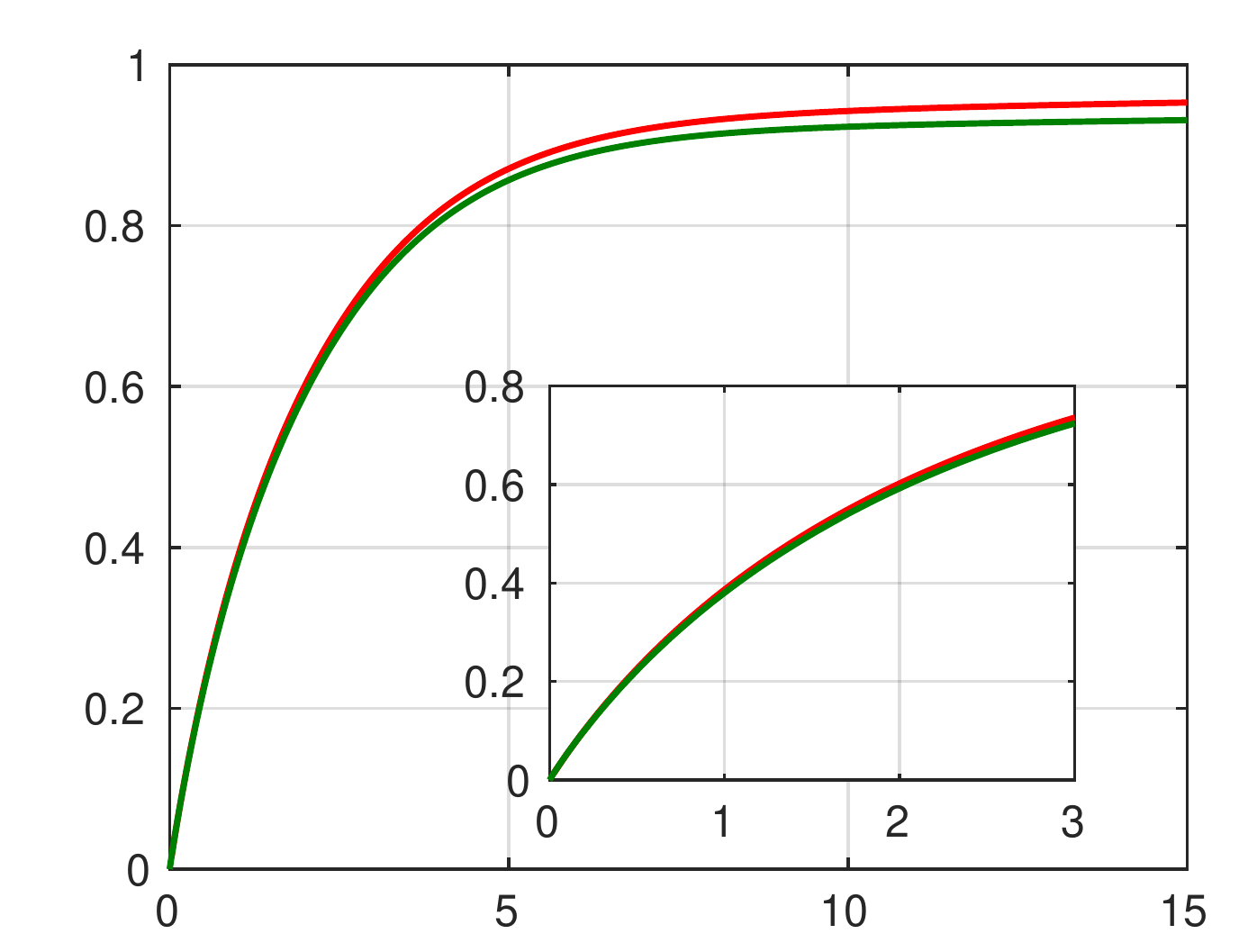}
  \end{minipage}
    \centering
    \begin{minipage}{2.6ex}
\rotatebox{90}{\hspace{2ex}{$\epsilon_{u}$}}
\end{minipage}
  \centering
  \begin{minipage}{0.46\linewidth}
  		\raggedright{(\textit{b})}
  		\includegraphics[width=\linewidth,trim={3.4cm 8.7cm 4cm 8.5cm},clip]
      {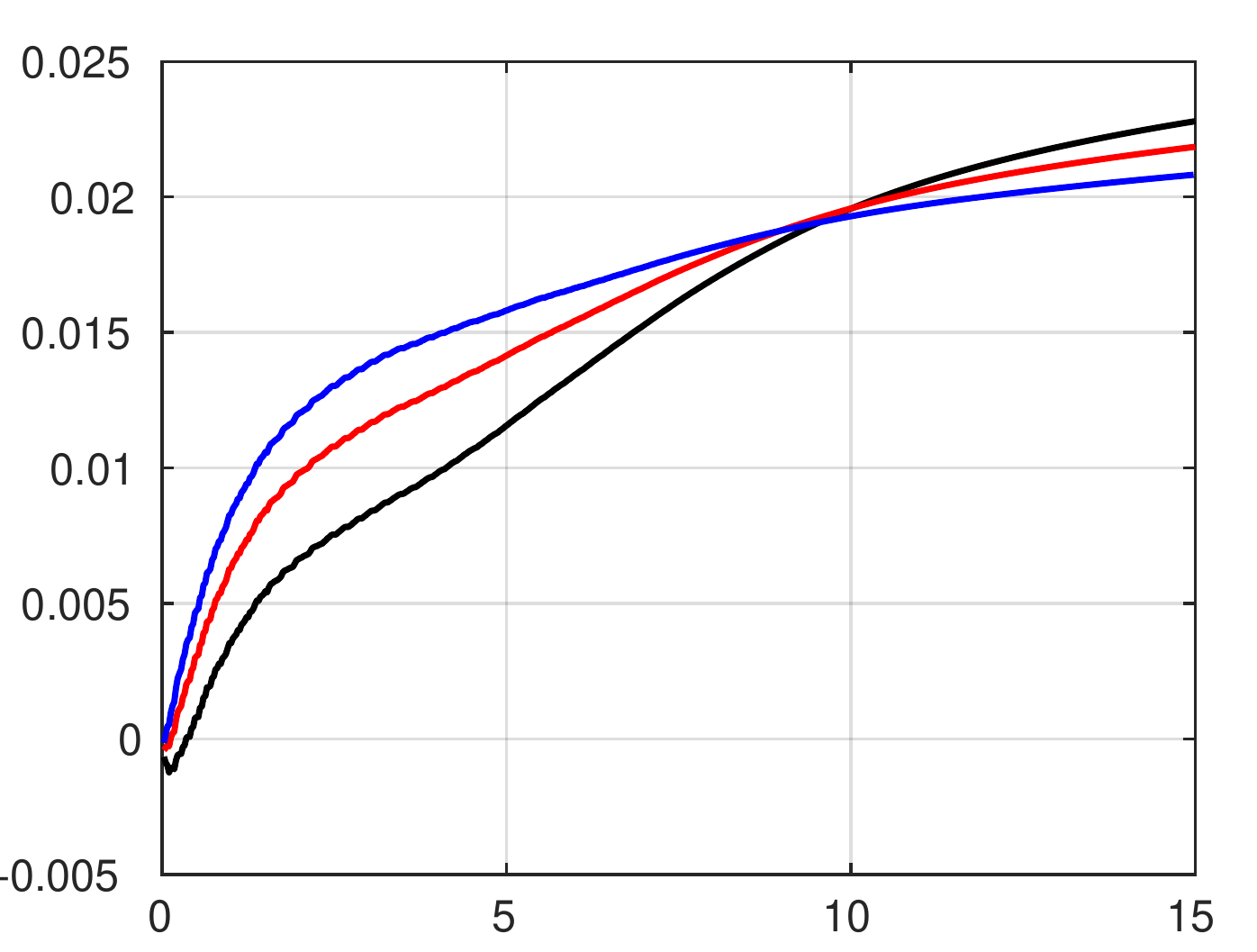}
  \end{minipage} \\
  \centering
    \begin{minipage}{2.6ex}
\rotatebox{90}{\hspace{2ex}{$v_p/U$}}
\end{minipage}
  \centering
  \begin{minipage}{0.46\linewidth}
      \raggedright{(\textit{c})}
      \includegraphics[width=\linewidth,trim={3.4cm 8.7cm 4cm 8.5cm},clip]
      {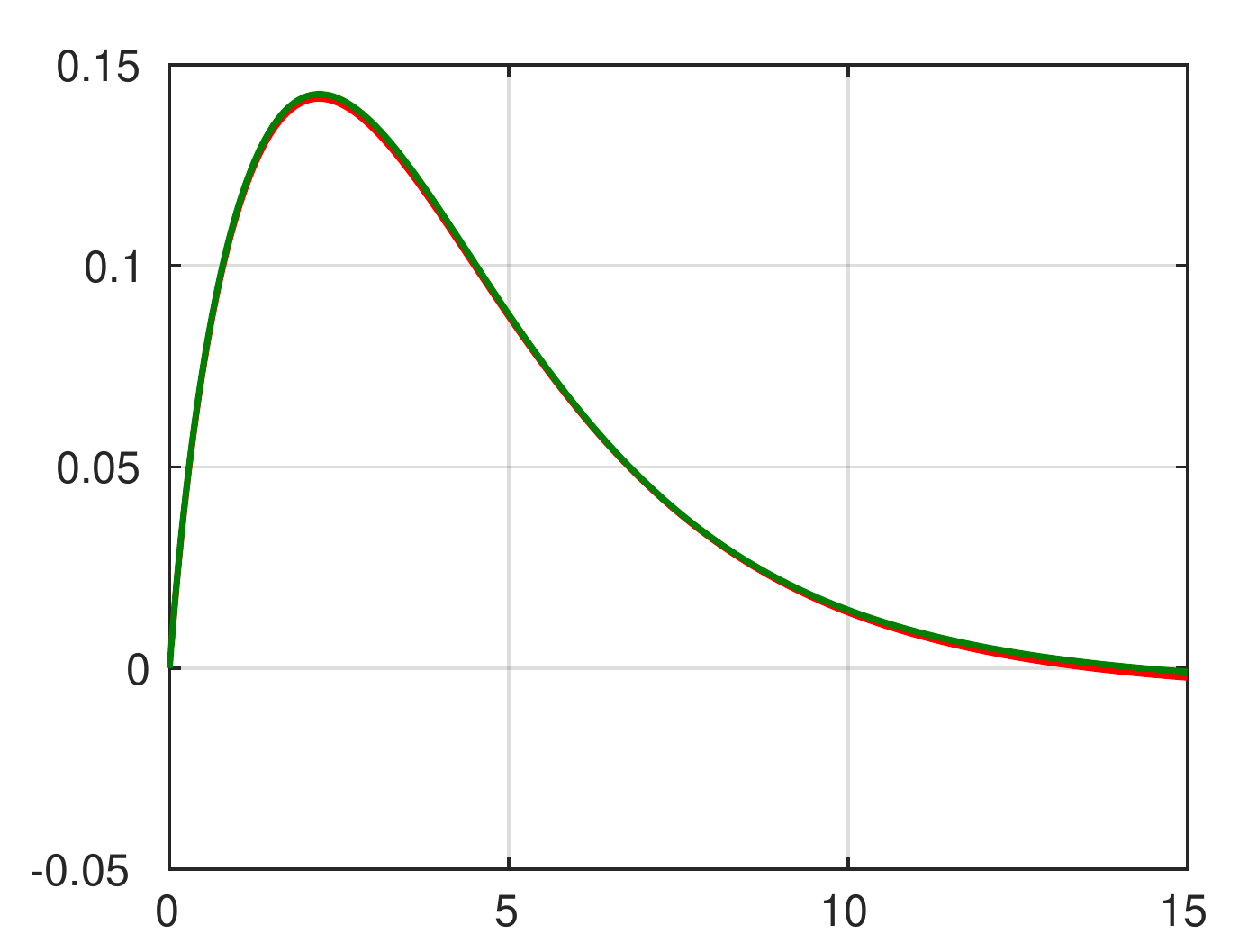}
  \end{minipage}
    \centering
    \begin{minipage}{2.6ex}
\rotatebox{90}{\hspace{2ex}{$\epsilon_{v}$}}
\end{minipage}
  \centering
  \begin{minipage}{0.46\linewidth}
  		\raggedright{(\textit{d})}
  		\includegraphics[width=\linewidth,trim={3.4cm 8.7cm 4cm 8.5cm},clip]
      {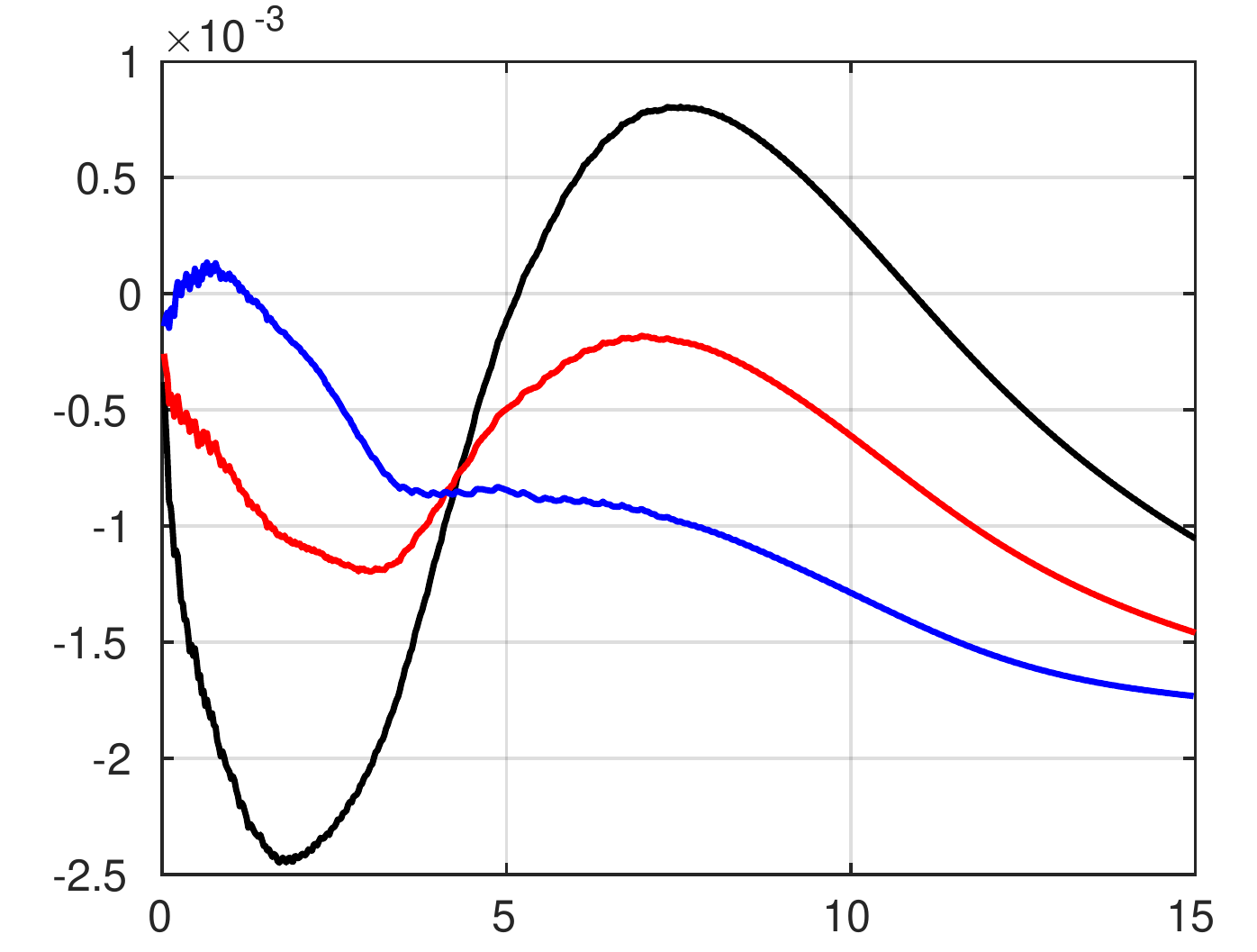}
  \end{minipage}\\
  \centering
    \begin{minipage}{2.6ex}
\rotatebox{90}{\hspace{2ex}{$-\omega_{pz} D/U$}}
\end{minipage}
  \centering
  \begin{minipage}{0.46\linewidth}
      \raggedright{(\textit{e})}
      \includegraphics[width=\linewidth,trim={3.4cm 8.7cm 4cm 8.5cm},clip]
      {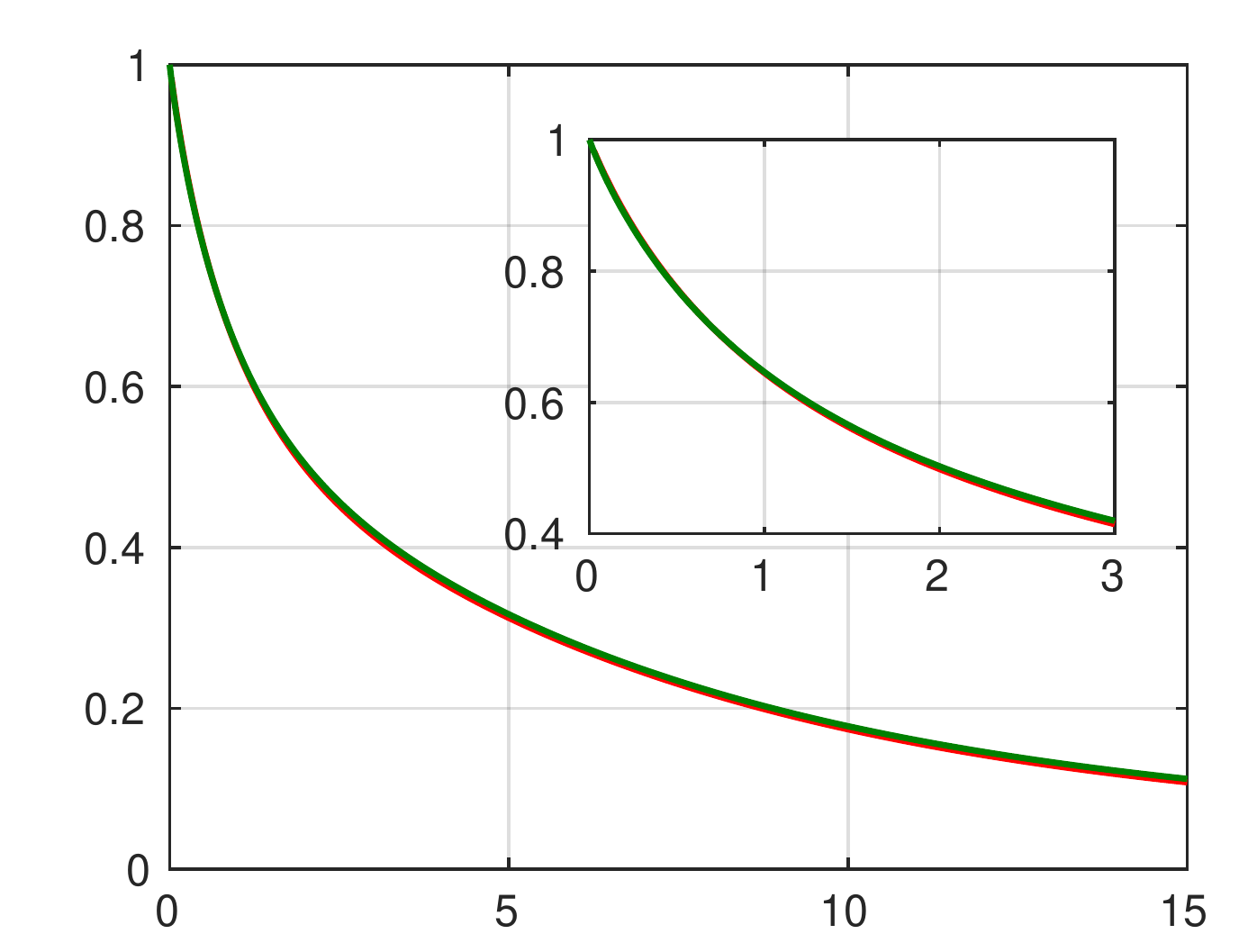}
              \centering{$t\, U/D$}
  \end{minipage}
    \centering
    \begin{minipage}{2.6ex}
\rotatebox{90}{\hspace{2ex}{$\epsilon_{\omega}$}}
\end{minipage}
  \centering
  \begin{minipage}{0.46\linewidth}
  		\raggedright{(\textit{f})}
  		\includegraphics[width=\linewidth,trim={3.4cm 8.7cm 4cm 8.5cm},clip]
      {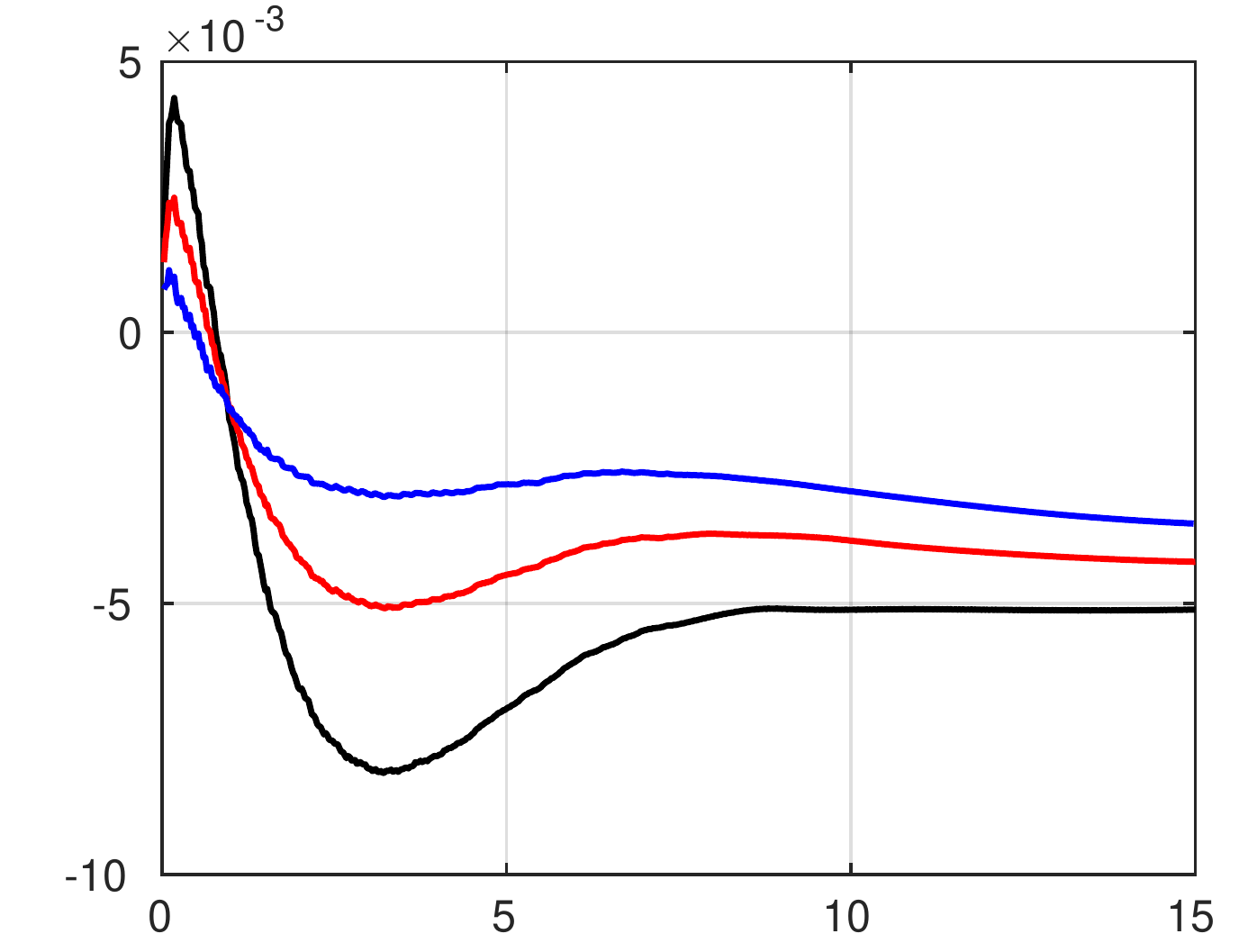}
              \centering{$t\, U/D$}      
  \end{minipage}
\caption
{Time evolution of 
(\textit{a}) particle streamwise velocity, $u_p/U$, 
(\textit{c}) particle lateral velocity, $v_p/U$,
(\textit{e}) particle angular velocity, $\omega_{pz}D/U$,
and error with respect to reference data of
(\textit{b}) streamwise velocity, $\epsilon_{u}$, 
(\textit{d}) lateral velocity, $\epsilon_{v}$,
(\textit{f}) angular velocity, $\epsilon_{\omega}$.
Green lines, reference data. 
$D/\Delta x=18$, Black.
$D/\Delta x=24$, Red.
$D/\Delta x=36$, Blue.
Insets in (\textit{a}) and (\textit{c}) highlight the initial phase.
\MGVN{In (\textit{a}), (\textit{c}), (\textit{e}) only the
results with $D/\Delta x=24$ are compared to the reference data.}
}
\label{fig:comp_dusek_nb_rot}
\end{figure}

Figure \ref{fig:comp_dusek_nb_rot}$(a,c,e)$ shows the time evolution of the
particle's streamwise and lateral velocity components and the
particle's angular velocity.
\MU{%
  The corresponding error evolutions according to the definition
  (\ref{equ-error-definition}) are shown in
  panels~\ref{fig:comp_dusek_nb_rot}$(b,d,f)$, where we use
  $\phi_{norm}=U$ for the lateral velocity and $\phi_{norm}=U/D$ for the angular
  velocity. 
}
The agreement with the reference data is generally very good for all
three quantities.  
The largest error amplitudes (smaller than 2.5\% at any time during
the simulation interval) are recorded for the streamwise
velocity. This is in line with our observations in the case of a
non-rotating sphere in a uniform flow at the same Reynolds number
(\S~\ref{sec:nb_free_stream}), with even the temporal evolution of the
errors featuring an analogous cross-over at some intermediate time,
and the finest grid yielding the smallest error at later times.
It can also be seen that the lateral linear particle velocity as well
as its angular velocity are already extremely well captured on the
coarsest grid.

\begin{figure}[ht]
  \centering
      \begin{minipage}{2.6ex}
\rotatebox{90}{\hspace{2ex}{$u_p/U$}}
\end{minipage}
  \begin{minipage}{0.46\linewidth}
      \raggedright{(\textit{a})}
      \includegraphics[width=\linewidth,trim={4.2cm 9cm 4cm 8.5cm},clip]
      {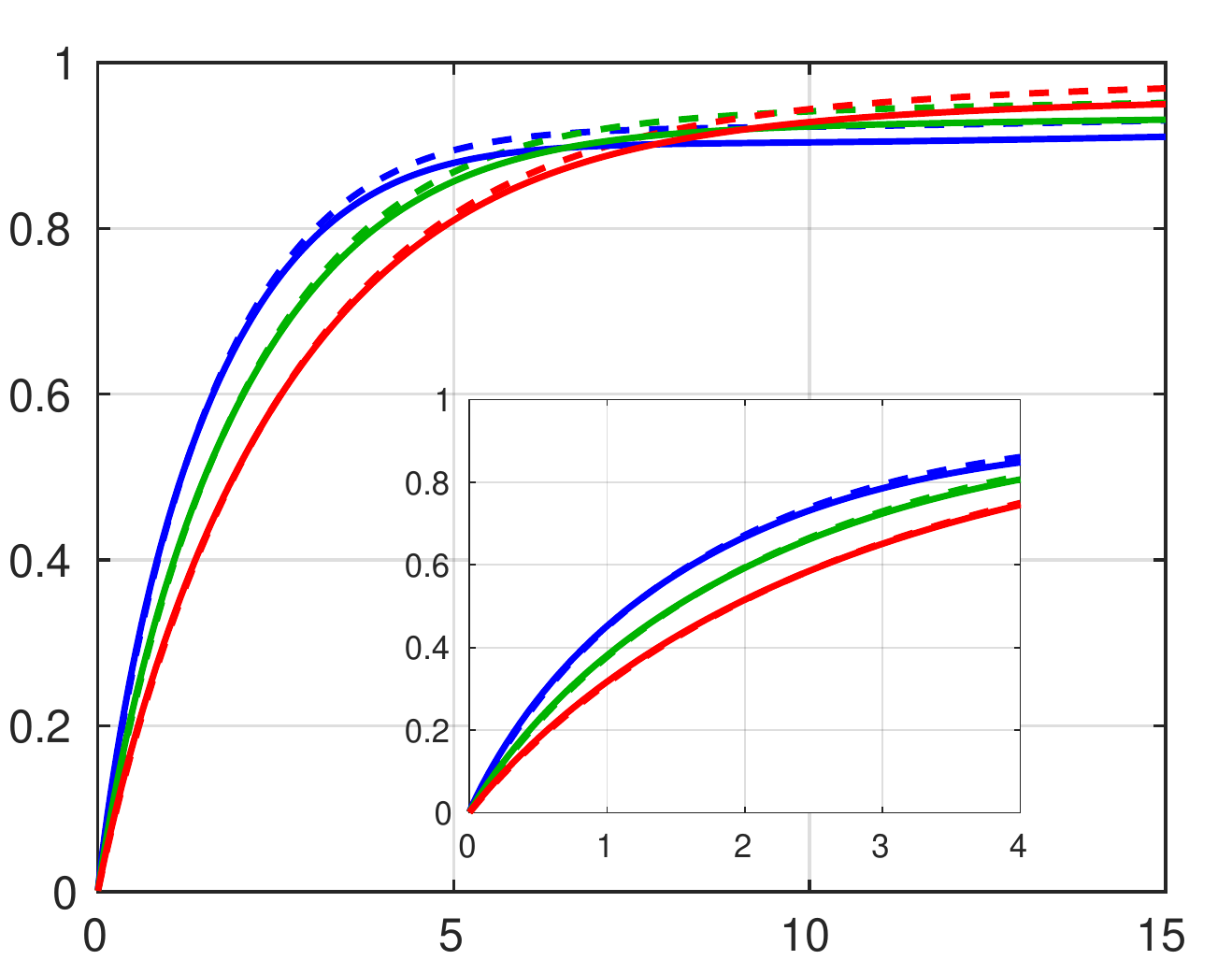}
           \centering{$t\, U/D$}
  \end{minipage}
  \centering
        \begin{minipage}{2.6ex}
\rotatebox{90}{\hspace{2ex}{$v_p/U$}}
\end{minipage}
  \begin{minipage}{0.46\linewidth}
  		\raggedright{(\textit{b})}
  		\includegraphics[width=\linewidth,trim={4.2cm 9cm 4cm 8.5cm},clip]
      {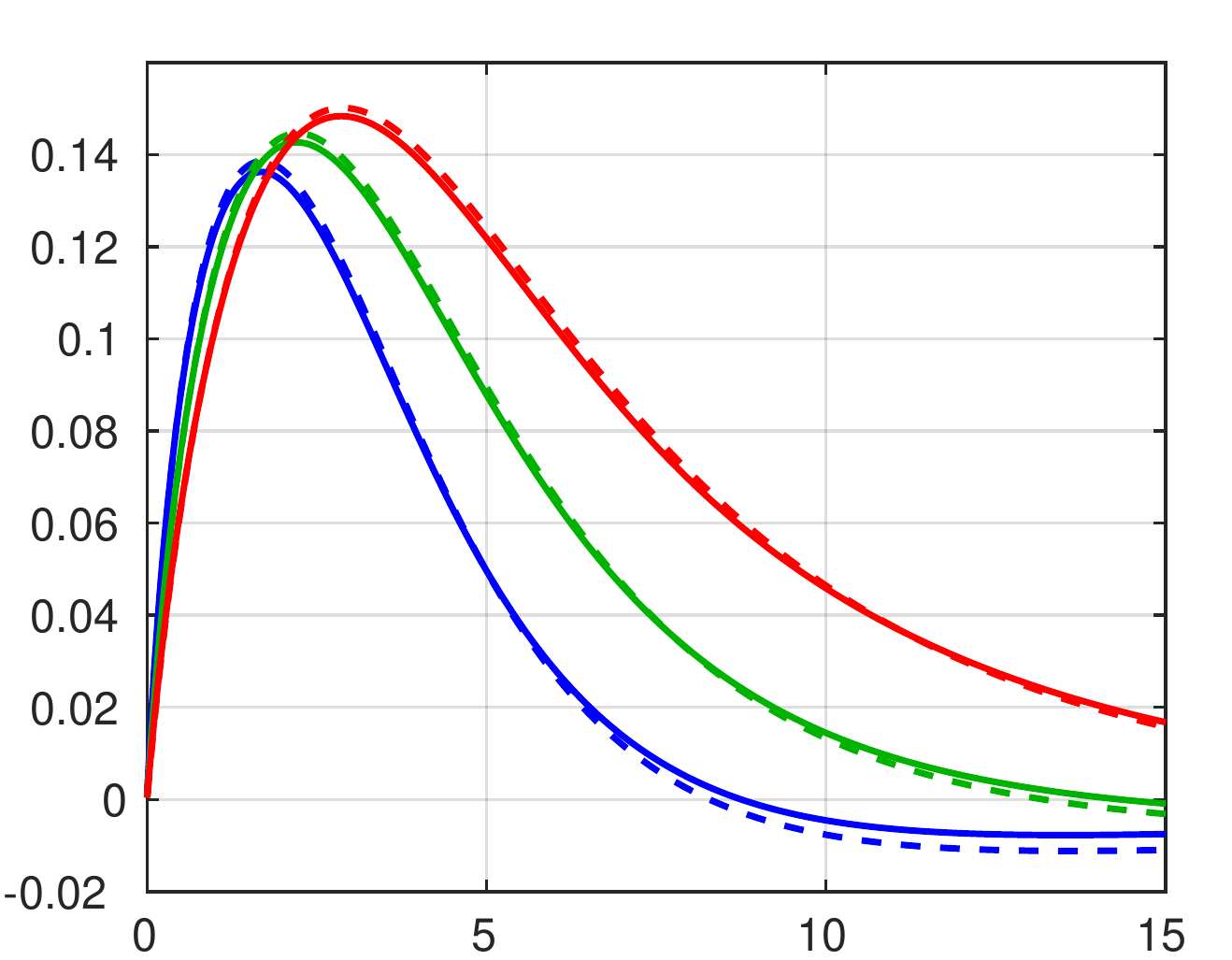}
           \centering{$t\, U/D$}
  \end{minipage}
   \centering 
       \begin{minipage}{2.6ex}
\rotatebox{90}{\hspace{2ex}{$-\omega_{pz} D/U$}}
\end{minipage}
  \begin{minipage}{0.46\linewidth}
      \raggedright{(\textit{c})}
      \includegraphics[width=\linewidth,trim={4.2cm 9cm 4cm 8.5cm},clip]
      {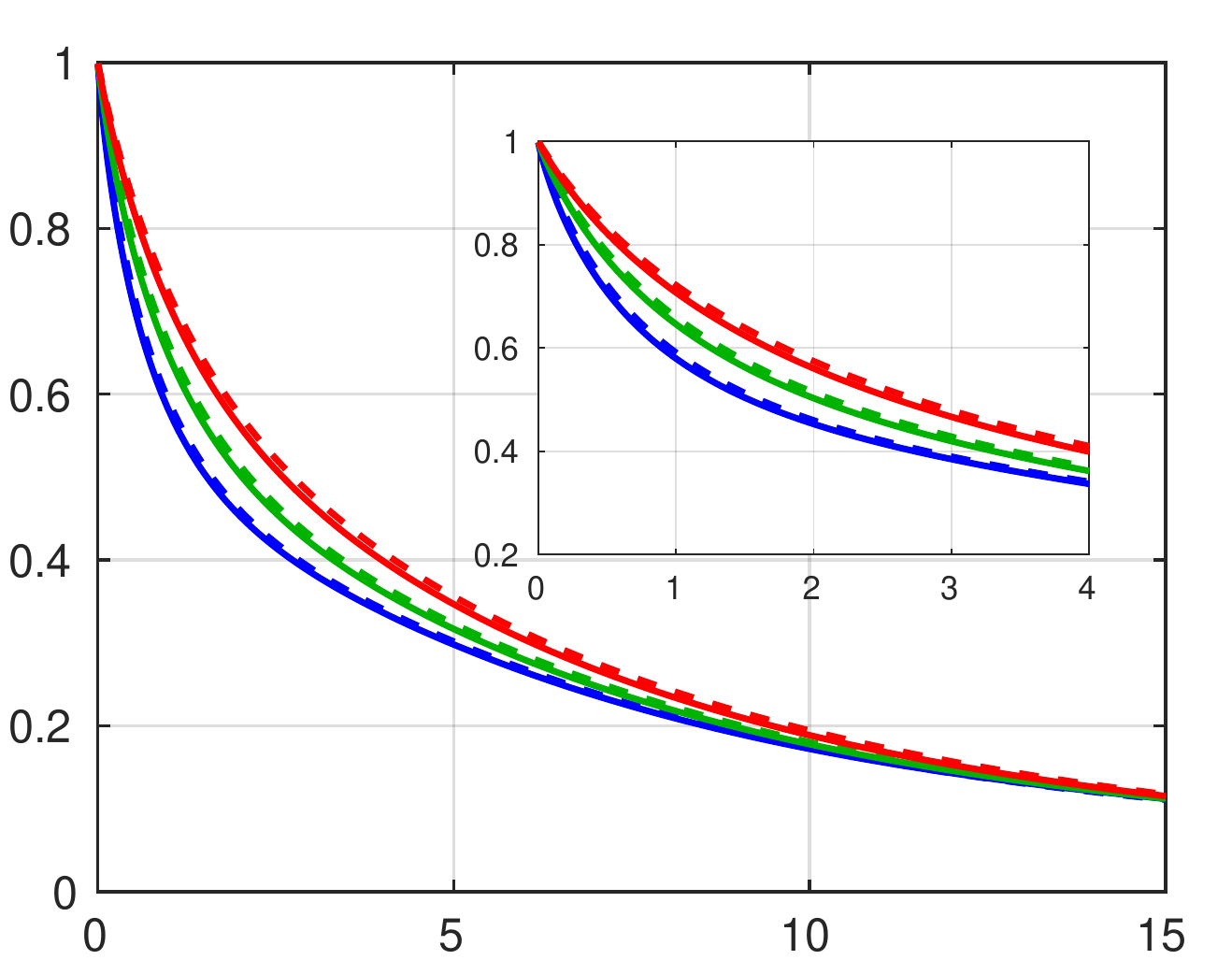}
           \centering{$t\, U/D$}
  \end{minipage}  
\caption
{Time evolution of 
(\textit{a}) particle streamwise velocity component, $u_{p}/U$,   
(\textit{b}) particle lateral velocity component, $v_{p}/U$, and 
(\textit{c}) particle spanwise angular velocity component, $\omega_{pz}D/U$.
Solid lines, reference data. 
Dashed lines, present results \MGV{with $D/\Delta x=24$}.
Blue, $\rhop/\rhof=0.6$.
Green, $\rhop/\rhof=1$.
Blue, $\rhop/\rhof=1.5$.
Insets in (\textit{a}) and (\textit{c}) highlight the initial phase.
}
\label{fig:comp_rot_rhop}
\end{figure}

Finally, we have performed analogous simulations for a 
particle with excess density ($\rho_p/\rho_f=1.5$) and for a less
dense particle ($\rho_p/\rho_f=0.6$) at a fixed numerical resolution
$D/\Delta x=24$. 
Figure \ref{fig:comp_rot_rhop} shows the time evolution of the
Lagrangian quantities (the particle's streamwise and lateral velocity
components and its particle's angular velocity) for the three density
ratios considered.  
The lower the density ratio the more rapid is the particle response
after its release, i.e.\ a steeper slope is encountered for short
times for all three of these quantities.
This effect is very well captured by the IBM simulations, which 
yield a similarly good agreement with respect to the reference
simulations irrespective of the value for the density ratio. 
This means that the present approach to treating density-matched
particles can also be applied with very good results to
non-density-matched particles. 

\section{Conclusions} \label{sec:concl}

In this work we have presented a simple modification of the
direct-forcing immersed boundary method (IBM) originally proposed by 
\citet{uhlmann:2005} in order to
enable it to be applied to particulate flows 
with solid-to-fluid
density ratios
around unity. 
%
Whereas the original formulation features a singularity for
neutrally buoyant mobile particles, 
the present approach is free from such a singularity. 
The idea is similar to the approach taken by \citet{tschisgale2017};
however, in the present case we apply the volume forcing term to the 
entire space occupied by the immersed solid object (instead of to the
vicinity of its interface only). This leads to a method which can be
used seamlessly for density ratios down to $\rho_p/\rho_f>0.5$. 
\revision{
As compared to previous approaches, no addition of stabilizing virtual
mass forces is required, and the method is free from spurious
oscillations. }
{As compared to previous approaches the method is free from spurious
oscillations. At the same time it is of a similar computational
complexity as the method of \citet{tschisgale2017} which requires an integral over the volume of the particle to be evaluated.}
%
%

The main difference with respect to the original formulation in
\cite{uhlmann:2005} lies in the particle velocity update which is
performed directly after the preliminary (composite) velocity field
has been computed in the absence of any IBM forcing terms. 
Then the subsequent computation of the volume force at the Lagrangian
positions already takes into account the updated velocity. One
additional ingredient which arises through this procedure is the
necessity to evaluate integrals of the linear and angular velocity
field over the volume occupied by the solid particle
\citep[similar to][]{kempe2012}, which can be evaluated efficiently at
second-order accuracy as sums over the respective quantities available
at the Lagrangian force points. 

%
%
%
%

In a first step, the new formulation has been validated using two flow
configurations which have been previously established in the
literature:
(i) lateral migration of a neutrally buoyant circular particle in
two-dimensional Couette flow for which reference data generated with a
fictitious domain method with Lagrange multipliers and a
finite-element discretization \citep{pan2013} is available; 
(ii) the release from rest of a neutrally buoyant sphere in a free
stream with reference data from spectral-element computations using a
body-conforming particle-attached mesh \citep{jenny:04b,tschisgale2017}. 
In the latter case, the resulting motion of the particle is
unidirectional without rotation. 
In order to subject the present method to a more demanding test, 
we have introduced a new configuration in which a particle in a free
stream is released after an initial phase in which it is
translationally fixed with an imposed constant angular velocity. The ensuing
unconstrained motion then features time-dependent streamwise and
lateral motion as well as rotational dynamics. 
For this case we have generated high-fidelity reference data with the
aid of the spectral-element method of \citep{jenny:04b} which can
henceforth be used for the cross-validation of numerical approaches.
%
We have tested an implementation of our new IBM formulation in a
fractional-step context, using a standard semi-implicit
Runge-Kutta method in conjunction with second-order
finite-differences on a staggered, uniform grid. 
In all three test cases the present formulation yields a very good
agreement with the available reference data. 
%
%
We conclude from our study that the proposed approach is a
cost-efficient and accurate modification of the original method which
allows for the simulation of fluid systems involving density-matched
solid particles.
%
One of the questions which we will address with the aid of the present
method in a future investigation is the occurrence of preferential
concentration in homogeneous-isotropic turbulence for finite-size,
neutrally buoyant particles \citep{fiabane:12}.
\revision{}{In a future communication we will consider the case of
neutrally-buoyant non-spherical particles. The present method appears 
well-suited to that case, as it can be applied with only minor
adjustments.}

\section*{Acknowledgments}

This work was supported by the German Research Foundation (DFG) under Project UH 242/11-1.
The work was partially performed during visits of MGV to the Karlsruhe Institute of Technology funded by the aforementioned project under the Mercator Fellow scheme.
BF was supported by a collaboration scholarship of the Spanish Ministry of Education.
We thankfully acknowledge the help of Dr. Gonzalo Arranz with the code implementation.
\revision{}{We also thank Prof. Tobias Kempe for providing his data in electronic
form.}


\appendix
\section{Description of reference computations with a spectral element method}
\label{app:spectral}

The numerical method utilized for the reference computations is
\MGV{identical} to the one employed in \citep{uhlmann2014}, as previously
developed in a series of papers \citep{ghidersa:00,jenny:04b,kotouc:08}. 
It utilizes a body-conforming mesh which translates 
with the particle's center of mass. 
The computational domain is of cylindrical shape with the
cylinder axis aligned in the vertical direction, a \MGV{radius} equal to
$8D$ and a height of $37D$ (with the particle center located at a
distance $12D$ from the cylinder base). 
The spatial discretization uses
truncated Fourier series in the azimuthal direction and a
spectral-element approach in the axial-radial plane (with
two-dimensional Legendre polynomial approximation inside the
elements). 
The particle motion is strongly coupled to the fluid
solver, as proposed in \citep{jenny:04b}, and a third-order
Adams-Bashforth method is used for the temporal
discretization. 
Ambient flow conditions are imposed at the cylinder
base, while a zero-stress condition is used at the upper cylinder
boundary and at the sides. Likewise, the pressure is set to zero on
the boundaries.
\MGV{The azimuthal complex Fourier expansion was truncated at the 7-th mode,}
and for all
elements \MGV{6} collocation points were used in each of the two
respective spatial directions. 
\MGV{
The element mesh was the same as that shown in figure 3b  of \citep{uhlmann2014} around the sphere but was extended to larger radius and height resulting in a larger number of elements (203).
}
The time step was adjusted such that the maximum $CFL$ number equals
$0.25$.
The accuracy of the reference simulations with this choice of
numerical parameters has been demonstrated through extensive
validation in previous publications \citep{jenny:04b,bouchet:06}. 

\bibliography{biblio}

\end{document}